\documentclass[journal=jacsat,manuscript=article]{achemsoz}
\usepackage[version=3]{mhchem} 
\usepackage{graphicx}
\usepackage{float}
\usepackage{subcaption}
\usepackage{color}
\usepackage{gensymb}
\usepackage{setspace}
\makeatletter
\author{Mohamed Zbiri}
\affiliation{Institut Laue-Langevin, 71 Avenue des Martyrs, Grenoble Cedex 9 38042, France.}
\email{zbiri@ill.fr}
\author{Peter A. Gilhooly-Finn}
\affiliation{Department of Chemistry, Queen Mary University of London, Mile End Road, London E1 4NS, United Kingdom}
\author{Peter Fouquet}
\affiliation{Institut Laue-Langevin, 71 Avenue des Martyrs, Grenoble Cedex 9 38042, France.}
\author{Christian B. Nielsen}
\affiliation{Department of Chemistry, Queen Mary University of London, Mile End Road, London E1 4NS, United Kingdom}
\author{Anne A. Y. Guilbert}
\affiliation{Department of Physics and Centre for Plastic Electronics, Imperial College London, Prince Consort Road, London SW7 2AZ, United Kingdom}
\email{a.guilbert09@imperial.ac.uk}
\makeatletter
\let\acs@address@list\relax
\makeatother
\title{Structural Dynamics of Polymer:Non-Fullerene Organic Solar Cell Blends: A Neutron Spectroscopy Perspective}
\begin{document}
\begin{abstract}
Organic solar cells (OSCs) based on ADA-type (acceptor-donor-acceptor) non-fullerene acceptors (NFAs) exhibit improved power conversion efficiency (PCE) compared to the conventional fullerene-based analogues. The optoelectronic properties of the OSC active layer are correlated to the underlying
structural dynamics of the active layer blend, and therefore influence the device performance. Using synergistically different neutron spectroscopy techniques, we studied the dynamics of binary and ternary blends made of the non-fullerene acceptors O-IDTBR and O-IDFBR, and the regio-regular donor polymer poly(3-hexylthiophene) (P3HT). Considering key factors like structural relaxation, miscibility and morphology, within a chosen temperature range, we probed the dynamical responses of the neat phases and the blends using time-of flight quasielastic neutron scattering and neutron spin echo measurements, complemented by neutron vibrational spectroscopy. We cover the femtosecond to nanosecond time window directly relevant to the operating active layer molecular processes. Blends of protonated O-IDTBR and O-IDFBR with either deuterated or protonated P3HT were considered for a contrast variation purpose. Our study confirms the observation of the hypo-miscibilty of O-IDTBR in the binary P3HT:O-IDTBR compared to O-IDFBR in P3HT:O-IDFBR and points towards a molecular alloying character of the NFA blend O-IDTBR:O-IDFBR, not observed in the ternary P3HT:O-IDTBR:O-IDFBR. A main outcome of this work is the evidenced similar dynamical response of the donor and acceptor components in both the binary and ternary blends, within the accessible experimental picosecond-nanosecond time window and up to 400 K. This finding is in contrast with our previous neutron spectroscopy and molecular dynamics studies of the fullerene-based blend P3HT:PCBM (PCBM: phenyl-C61-butyric acid methyl ester), where we highlighted distinct behaviors of P3HT and PCBM in the blend in terms of the vitrification/frustration of P3HT upon blending and the plasticization of PCBM by P3HT. Alike P3HT vitrification is not presently observed, which we ascribe to the resemblance of the chemical structures of O-IDTBR/O-IDFBR, and P3HT. The absence or the weak vitrification evidenced here is in line with recent reports and is likely related to the improved PCE exhibited by the ADA-type NFA-based OSCs.
\end{abstract}
\section{Introduction}
\begin{figure}[H]
\includegraphics[width=0.75\textwidth]{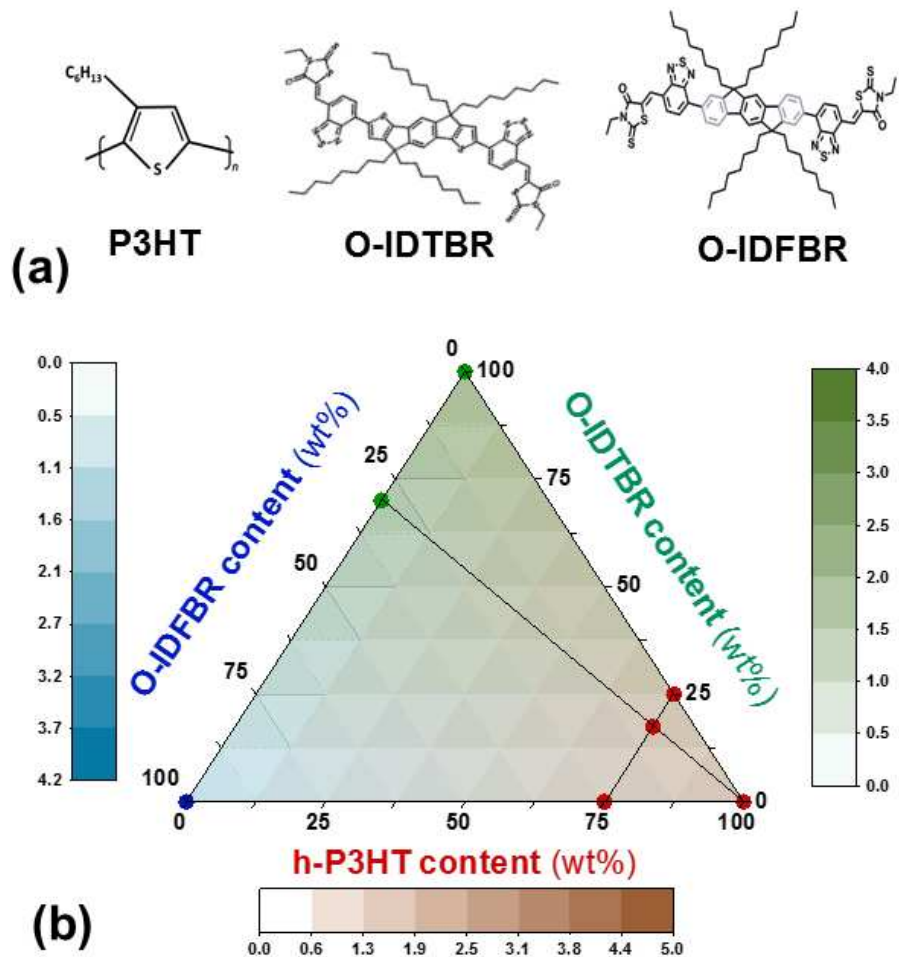}
\caption{(a) Illustration of the chemical structures of the donor polymer P3HT (left), and the non-fullerene acceptor molecules O-IDTBR (middle) and O-IDFBR (right), forming the ternary donor:acceptor:acceptor, binaries donor:acceptor or binary acceptor:acceptor. The light grey color in O-IDFBR is used to mark the structural difference with O-IDTBR, namely benzenes are replaced by thiophenes. The two molecules resemble chemically each other, except that O-IDTBR is geometrically planar compared to the twisted O-IDFBR, which plays an important role in their degree of crystallinity in the blend. (b) Color-coded representation of the chemical composition (wt\%) and the incoherent neutron cross-sections $\sigma$ (cm$^{-1}$) of the ternary blend h-P3HT:O-IDTBR:O-IDFBR and the related three binary blend configurations, along with their neat constituents, where h-P3HT denotes the protonated polymer in contrast to the deuterated case also considered in this work and labelled in the text as d-P3HT. Values of the studied compositions and the corresponding $\sigma$ are gathered in Table S1 (supplementary information). The blue point represents the neat O-IDFBR sample. The two green points represent the neat O-IDTBR and the binary blend O-IDTBR:O-IDFBR, which is dominated by the O-IDTBR signal. The four red points represent the neat h-P3HT, the two blends h-P3HT:O-IDTBR and h-P3HT:O-IDFBR, and the ternary blend h-P3HT:O-IDTBR:O-IDFBR, which are all dominated by the h-P3HT signal.} 
\label{fig:fig1}
\end{figure}

The active layer of polymeric organic solar cells (OSCs) is composed of an electron-donating polymer and (an) electron-accepting small-molecule semiconductor(s). The electron-accepting organic semiconductors, which have been developed to date, can be broadly separated into two classes: (i) fullerene acceptors and (ii) non-fullerene acceptors (NFAs). Although NFAs were developed early on for vacuum-deposited OSCs, fullerene acceptors have been dominating the field since the introduction of the first solution-processed bulk heterojunction due to their compatibility with a large range of donor materials and relative ease of processing~\cite{Armin2021}. Early NFAs were suffering from a strong tendency to aggregate due to strong $\pi-\pi$ interactions stemming  partially from their 2D planar shape. Thus, large domains and aggregates of acceptors were forming in the active layer leading to poor efficiencies in comparison with fullerene acceptors. Fullerene acceptors are, however, poorly harvesting the solar energy. Recently, ADA-type (Acceptor-Donor-Acceptor) NFAs started to compete with fullerene acceptors~\cite{Naveed2018}. ADA-type NFAs exhibit three structural elements: the core, the end units and the side chains enabling controlling the energy levels, optical properties, solubility, crystallinity and electron mobility. In particular, the excessive aggregation observed in earlier NFAs could be controlled by using twisted cores leading to "3D-like" molecules and by engineering bulky "out-of-plane" side chains. The acceptor groups (end units) can be chosen to ensure good $\pi-\pi$ stacking. Recent efforts have made it possible to synthesize families of NFAs via well-established molecular-by-design routes, aiming at reaching a power conversion efficiency (PCE) milestone as high as 20\%~\cite{Armin2021,Gillett2021,Gao2022}.
\\
Simplistically, mixed phases are beneficial for efficient charge separation, while pure interconnected domains are beneficial for charge transport. Donor and acceptor materials are partially miscible~\cite{Naveed2018, Zbiri2021}. In polymeric solar cells, the asymmetry of molecular weights between the polymer donor and the small-molecule acceptor is reflected in the asymmetry of the miscibility gap and leads to the coexistence of polymer-rich domains, relatively pure small-molecule acceptor domains and polymer aggregates or crystallites when the conjugated polymer is semi-crystalline. Eutectic phase diagrams have been reported for a number of systems, including both fullerene and NFAs~\cite{Muller2008, Wadsworth2018, Rezasoltani2020}. The composition of the active layer of the best-performing solar cells is often correlated with the eutectic composition, slightly richer in small-molecule acceptors to ensure sufficient connection between the nearly pure acceptor domains~\cite{Muller2008, Guilbert2012, Wadsworth2018, Rezasoltani2020}. For most material systems studied to date, the ideal content of small-molecules in the polymer-rich phase has been evaluated around 20-30\%~\cite{Ghasemi2019, Zbiri2021}, which corresponds to a trade-off between efficient charge separation and electron percolation to ensure the necessary electron transport~\cite{Saladina2021}. Within the Flory-Huggins theory framework, a polymer-rich phase containing around 20-30\% of small-molecules corresponds to a narrow window of ideal (amorphous-amorphous) Flory-Huggins parameters ($\chi_{aa}$) for the donor polymer:small-molecule acceptor system. 
\\
Not only is finding the right combination of materials challenging but also finding the right quench. If the system is hyper-miscible, the quench depth is too low and the binodal composition of the polymer-rich phase will contain too many small-molecule acceptors, which is usually detrimental for the charge separation process. If the system is hypo-miscible, the quench depth is too high and the thermodynamically-stable binodal composition of the polymer-rich phase will be too poor in small-molecule acceptors to reach the percolation threshold~\cite{Ghasemi2019,Ghasemi2021}. Below the glass transition of the system, thermodynamical relaxation of the polymer-rich phase to the binodal composition can still occur but is kinetically hindered. Hence, hypo-miscible systems can be kinetically quenched to bring the composition of the polymer-rich phase closer to the percolation threshold. Furthermore, small-molecule acceptors are usually crystalline. Nucleation, growth and coarsening of the crystalline small-molecule acceptor domains often occur over time depleting the surrounding regions of small-molecule acceptors. The composition of the depleted region is set by the composition at the liquidus of the system, which is lower than the percolation threshold. 
\\
Although kinetically trapping a favourable microstructure seems a good strategy, it is in practice challenging and the time evolution of such a microstructure can be detrimental to the lifetime of the OSCs. Introducing a second acceptor with a high-enough miscibility with the donor polymer to reach electron percolation has proven to be an efficient strategy to stabilise the morphology~\cite{Baran2017}.
\\
Beyond thermodynamic drivers, both the relaxation of the blends and the nucleation and growth of crystals are impacted by kinetic factors such as the diffusion of the small molecules within the polymer matrix. Hence, introducing a less ductile polymer can help, for example,  to stabilize the blend kinetically. Ghasemi \textit{et al.} measured the activation energies of the diffusion process of various NFAs in the polymer donor poly(3-hexylthiophene) (P3HT). The measured activation energies ranged from 30 to about 45 kcal.mol$^{-1}$. Those activation energies are significantly larger than the activation energy of 13 kcal.mol$^{-1}$ measured  for the diffusion of fullerene derivative [6,6]-Phenyl C61 butyric acid methyl ester (PCBM) in P3HT, with 13 kcal.mol$^{-1}$ is significantly higher than the thermal energy~\cite{Ghasemi2021}. The energy activation is likely linked with the energy necessary to create an additional void of the size of the small-molecule acceptors in order for the small-molecule to move. Diffusion processes are known to be impacted by molecular structure, size and shape. PCBM in P3HT has also a higher diffusion coefficient than the NFAs in P3HT. The high activation barrier for the NFAs to diffuse coupled with their low diffusion coefficients could be beneficial to achieve stability in NFA-based OSCs.
\\
Within the families of ADA-type NFAs, O-IDTBR ~\cite{Chen2017,Gasparini2017,Liang2018,Hoefler2018,Wadsworth2018,Corzo2019,Lopez-Vicente2021,Yan2021} and O-IDFBR~\cite{Baran2017,Rezasoltani2020,Yan2021} are representative materials and therefore, good material models. The structures of O-IDTBR and O-IDFBR are illustrated in Figure~\ref{fig:fig1}. O-IDTBR and O-IDFBR are structurally similar. O-IDTBR has a more planar structure than O-IDFBR, thus, retaining its degree of crystallinity in blends, making a potential difference in terms of their dynamical interaction with P3HT~\cite{Rezasoltani2020}. On one hand, O-IDTBR exhibits an eutectic phase diagram with P3HT with an optimum composition in terms of PCE close to the eutectic composition similarly to what was found for fullerene acceptors. On the other hand, O-IDFBR is found to mix better with P3HT, hindering P3HT crystallinity. The phase behavior of O-IDFBR differs with a high NFA content composition-window where the material is mainly amorphous. The optimum composition for P3HT:O-IDFBR is found to be at a composition where P3HT still retains its crystallinity and where the electron percolation between the nearly pure O-IDFBR domains is sufficient. Thus, P3HT:O-IDFBR-based solar cells exhibit significantly lower PCE than P3HT:O-IDTBR-based solar cells.\cite{Rezasoltani2020} Interestingly, it was shown that adding O-IDFBR, as a second acceptor and third component, in the binary blend P3HT:O-IDTBR can lead to a better PCE and an increased stability of the device~\cite{Baran2017}.
\\
Neutron spectroscopy offers master tools for probing the structural dynamics of organic semiconductors. Although it is still an under-used method in this field~\cite{Cavaye2019}, recent studies have demonstrated its power in the framework of (i) OSCs~\cite{Guilbert2012,Guilbert2016,Guilbert2017,Zbiri2021}, (ii) charge transport~\cite{Guilbert2015,Guilbert2019,Guilbert2021PTTPFS,Stoeckel2021} and (iii) photochemical water splitting materials~\cite{Guilbert2021CMP,Sprick2019,Zbiri2021CTF}.
Neutron spectroscopy encompasses techniques covering synergistically different time scales at the nanometer length scale. It can be used to gain deeper and newer insights into the structure-dynamics-properties relationship. In particular, time-of-flight (TOF) quasielastic neutron scattering (QENS) and neutron spin echo (NSE) methods cover together a picosecond-to-nanosecond time window. The time domain can be further extended to shorter time scales using neutron vibrational spectroscopy, dealing with vibrational modes in the femtosecond range. These time scales match perfectly those of the molecular electronic processes, e.g. charge hopping, in the active layers of OSCs. Furthermore, neutrons probe efficiently light elements such as hydrogens, unlike X-rays. Using the deuteration technique offers the unique opportunity of varying efficiently the contrast between different parts of the systems (for instance deuterating the donor in a blend to better highlight the acceptor structural dynamics in the blend). Neutrons do not interfere with the electronic behavior of materials~\bibnote{This is in contrast with optical spectroscopy, which is subjected to photoluminescence when dealing with conjugated organic materials. The light probe interacts with the optical and electronic processes of these systems. In this context, the absorption range of the sample and/or the measured signal contaminated with a contribution from subsequent light emission of the material can limit or make it difficult to use optical spectroscopy to study dynamics of organic semiconductors.}, allowing probing the full Brillouin zone without selection rules. All vibrational modes can hence be mapped out.
\\
We previously studied structural dynamics of the fullerene-based blend P3HT:PCBM using TOF-based QENS~\cite{Guilbert2016} and molecular dynamics (MD) simulations~\cite{Guilbert2017}. We evidenced opposite behaviors of P3HT and PCBM in the blend i.e. the vitrification/frustration of P3HT by PCBM and the plasticization of PCBM by P3HT~\cite{Guilbert2016}. Our MD analysis suggested that a partial wrapping of P3HT around PCBM was responsible for the experimentally-observed distinct behaviours~\cite{Guilbert2017}.
\\
In this paper, we study the dynamics of binary and ternary blends of O-IDTBR and O-IDFBR, with the regio-regular donor polymer P3HT (Figure \ref{fig:fig1}), using TOF-based QENS on a 5-50 picosecond time window. The QENS measurements are extended to the nanosecond timescale via NSE measurements. Neutron diffraction patterns extracted from the TOF measurements, as well as inelastic neutron scattering (INS) in terms of vibrational dynamics are considered to evaluate the microstructure and morphology. It is worth mentioning that diffraction patterns are neither a goal of neutron spectroscopy (QENS \& INS), nor presently intended for a quantitative purpose, but are rather shown to make qualitative observations and to strengthen, when relevant, the main spectroscopic interpretations. The outcome of this work is threefold: (i) regarding the morphology aspect, this study supports the higher miscibility of O-IDFBR with P3HT with respect to O-IDTBR. This study thus confirms the hypo-miscibility of the latter and its adequacy as a better candidate than O-IDFBR in binary OSCs~\cite{Rezasoltani2020,Gao2022}, (ii) our neutron measurements point towards an alloying character of the NFA blend O-IDTBR:O-IDFBR, which is not observed in the ternary blend P3HT:O-IDTBR:O-IDFBR, and (iii) interestingly, regarding the structural dynamics aspect, we find that unlike in our above mentioned case study of the fullerene-based blend P3HT:PCBM, P3HT dynamics is presently not impacted in the binary and ternary blends, by either NFA, O-IDTBR and O-IDFBR, within the accessible experimental picosecond-nanosecond time window, and up to 400 K. The absence or weakness of vitrification/frustration in the studied NFAs-based blends could be due to a resemblance of the chemical structures of O-IDTBR/O-IDFBR, and P3HT, leading to a dynamical behavior reflecting flexibility of motions of the blend components for an optimum electronic interaction. The highlighted vitrification-free aspect of NFA-based OSCs could hence explain their improved PCE comparing to the fullerene-based analogues. In this context, it is worth noticing that it has been shown recently that blends with hypo-miscibility and a weak vitrification are excellent candidates for efficient solar cells~\cite{Gao2022}. 
\\
\section{Results and discussion}
\subsection{Structural dynamics of the NFA molecules O-IDTBR and O-IDFBR}
\begin{figure}
\includegraphics[width=\textwidth]{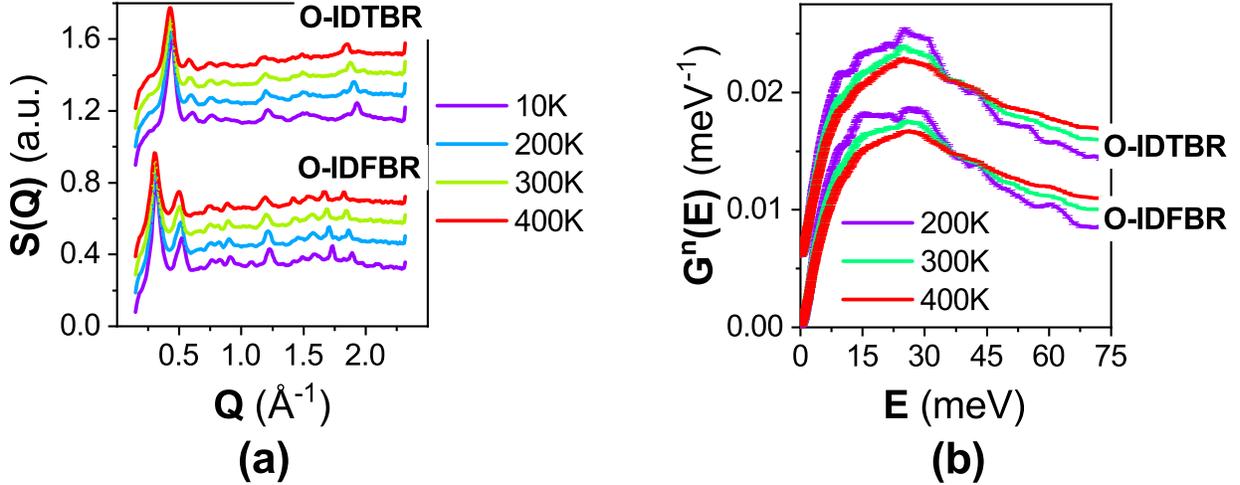}
\caption{Temperature evolution of (a) diffraction patterns and (b) generalized density of states (G$^{(n)}$(E)) of the non-fullerene acceptor molecules O-IDTBR and O-IDFBR, from neutron TOF-based measurements using the IN5 spectrometer, operating at an incident wavelength of 5\AA ~(E$_i$=3.27 meV). Spectra are vertically shifted for clarity.}
\label{fig:fig2}
\end{figure}
Neutron TOF-based spectroscopy allows to probe concomitantly the elastic, quasielastic and inelastic processes. We first characterize the structure and morphology of the NFA molecules, O-IDTBR and O-IDFBR, by considering the elastic and inelastic components to extract diffraction patterns and generalized density of states (GDOS), respectively, as shown in Figure \ref{fig:fig2}. Figure \ref{fig:fig2} (a) shows clear distinct Bragg peaks observed at low momentum transfers (low Q values) for both O-IDTBR and O-IDFBR, as well as less intense peaks at higher Q values. O-IDTBR presents a strong peak around 0.45 \AA$^{-1}$. On the other hand, in this low Q-region, O-IDFBR exhibits a split Bragg feature (two peaks around 0.4 and 0.5 \AA$^{-1}$) compared to O-IDTBR. At higher Q values, diffraction patterns of O-IDTBR show a peak around 1.9 \AA$^{-1}$ corresponding to the crystallographic plane (402) separating two O-IDTBR molecules in a $\pi-\pi$ stacking configuration. This leads to an interplanar distance of $\sim$ 3.27 \AA, matching well the reported range of 3.12 - 3.32 \AA ~\cite{Bristow2019,Halaby2021}. Assuming both NFAs have a very closely similar crystal structure (Figure~\ref{fig:fig1}(a)), the $\pi-\pi$ stacking in O-IDFBR is subjected to a larger separation than in O-IDTBR, as also reflected in the corresponding lower Q value, $\sim$ 1.8 \AA$^{-1}$, compared to IDTBR ($\sim$ 1.9 \AA$^{-1}$). This is likely due to the more twisted nature of the O-IDFBR core compared to the geometry of O-IDTBR. It is worth noticing the fact that these NFAs are not planar, it makes it delicate to derive precise intermolecular distances~\cite{Bristow2019}. The observed Bragg features shift slightly to lower Qs as temperature increases, reflecting the expected thermal expansion behavior. No phase transition is observed within the probed temperature range 10 - 400 K (no peak splitting, appearance or disappearance of peaks).\\
The similarity of the chemical structures of O-IDTBR and O-IDFBR is also reflected in their GDOS spectra, resembling closely each other (Figure \ref{fig:fig2} (b)). Subtle differences are noticed upon cooling, stemming from the reduced temperature-induced broadening due to the Debye-Waller effect. The two NFAs exhibit a similar Debye growth in the 0-10 meV energy range upon heating. This points towards a closely related morphological behavior, which is also confirmed by their similar vibrational features at higher energy transfer. Upon cooling distinct vibrational bands are observed in the 10 - 40 meV energy range. A vibrational band is observed at 10 meV, followed by two bands at 10-25 meV and 25-40 meV. The first band is stronger for O-IDFBR, while the second band is stronger for O-IDTBR. The subtle differences observed upon cooling between O-IDFBR and O-IDTBR in terms of their respective band intensities could be due to their geometrical difference and therefore their frozen-like packing and associated vibrational behavior. \\
\\
\begin{figure}[H]
\includegraphics[width=\textwidth]{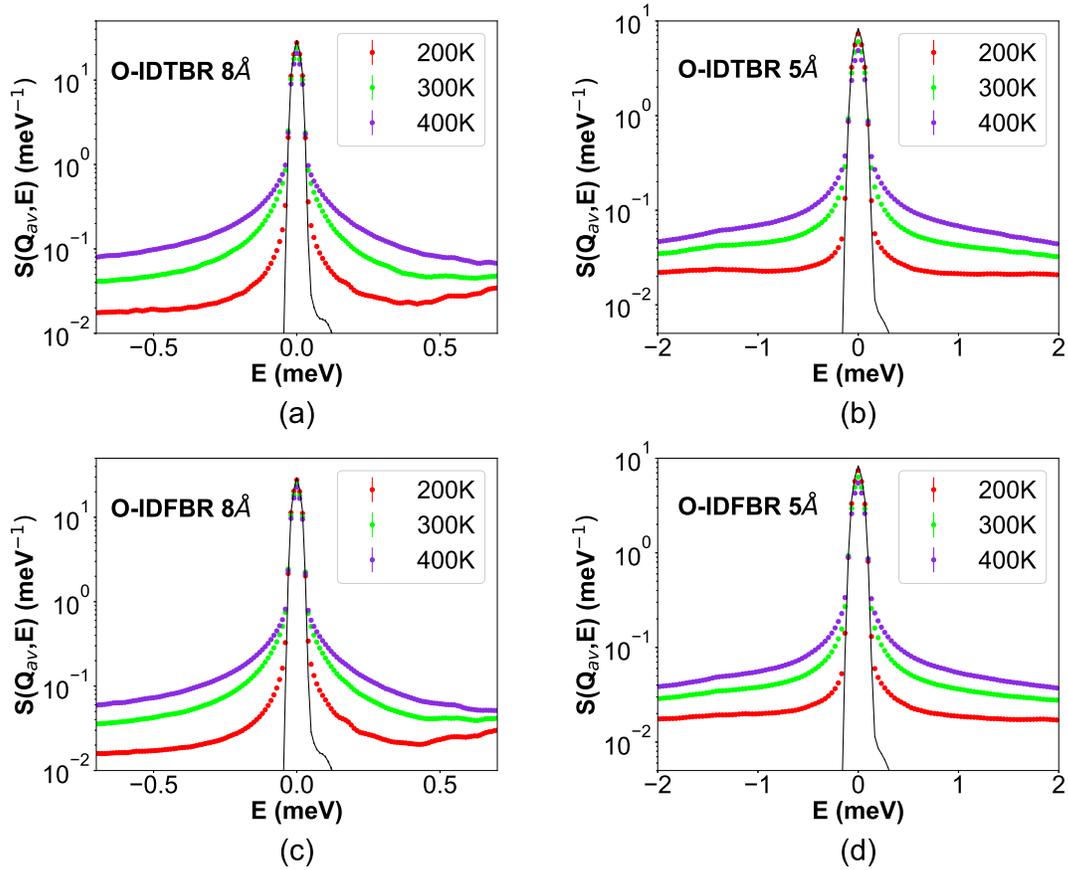}
\caption{The Q-averaged TOF-based QENS spectra from IN5 measurements, for the non-fullerene acceptor molecules O-IDTBR (a,b) and O-IDFBR  (c,d), using two neutron incident wavelengths of (a,c) 8 \AA ~($E_i$=1.28 meV) and (b,d) 5 \AA ~($E_i$=3.27 meV). The black line represents the instrumental resolution.}
\label{fig:fig3}
\end{figure}
We pursued the characterization of the dynamics of the two NFAs as function of temperature by TOF-based QENS technique to probe their local dynamical aspects (Figure \ref{fig:fig3}). Both NFAs exhibit a similar temperature-dependence behavior with a broadening of the QENS spectra upon heating, indicating an acceleration of the dynamics. However, the QENS spectra of O-IDFBR are systematically slightly narrower than O-IDTBR indicating that in the measured energy transfer range, structural dynamics of the twisted O-IDFBR are slightly slower than those of the planar O-IDTBR. Given the corresponding experimental TOF-related picosecond time window and the temperature range up to 400 K, we are likely observing here primarily the dynamics of the side chains of the NFAs. This slight reduction of O-IDFBR side chains motion in comparison with O-IDTBR is likely related to the twisted core of the O-IDFBR that induces a steric hindrance, hence constraining the local dynamics of the side chains. We went further in probing relaxational dynamics by extending the TOF-based picosecond timescale to cover the nanosecond region by carrying out NSE measurements (Figure \ref{fig:fig4}). NSE technique measures the incoherent intermediate scattering function $S_{inc}(\mathbf{Q},t)$, which is linked to the self-part of the van Hove function $G_s(\mathbf{r}, t)$ by a spatial Fourier transform:
\begin{equation}
    S_{inc}(\mathbf{Q},t) = \int d\mathbf{r} G_s(\mathbf{r}, t)e^{-i\mathbf{Q}.\mathbf{r}}
    \label{eq:FFT_to_intermediate}
\end{equation}
The QENS technique measures the dynamical structure factor, $S(\mathbf{Q}, \omega)$, convoluted with an instrumental resolution $R(\mathbf{Q}, \omega)$. For the samples studied here, the incoherent neutron cross-section is much larger than the coherent neutron cross-section and thus, $S(\mathbf{Q}, \omega)$ can be well approximated in terms of its incoherent component $S_{inc}(\mathbf{Q}, \omega)$. $S_{inc}(\mathbf{Q}, \omega)$, from QENS, and $S_{inc}(\mathbf{Q},t)$, from NSE, are related to each other by a Fourier transform as:
\begin{equation}
    S_{inc}(\mathbf{Q},\omega) \approx \int dt S_{inc}(\mathbf{Q}, t)e^{-i\omega t}
\end{equation}
Or via the inverse Fourier transform as:
\begin{equation}
    S_{inc}(\mathbf{Q},t) \approx \int dt S_{inc}(\mathbf{Q}, \omega)e^{i\omega t}
\end{equation}
\begin{figure}[!h]
\includegraphics[width=\textwidth]{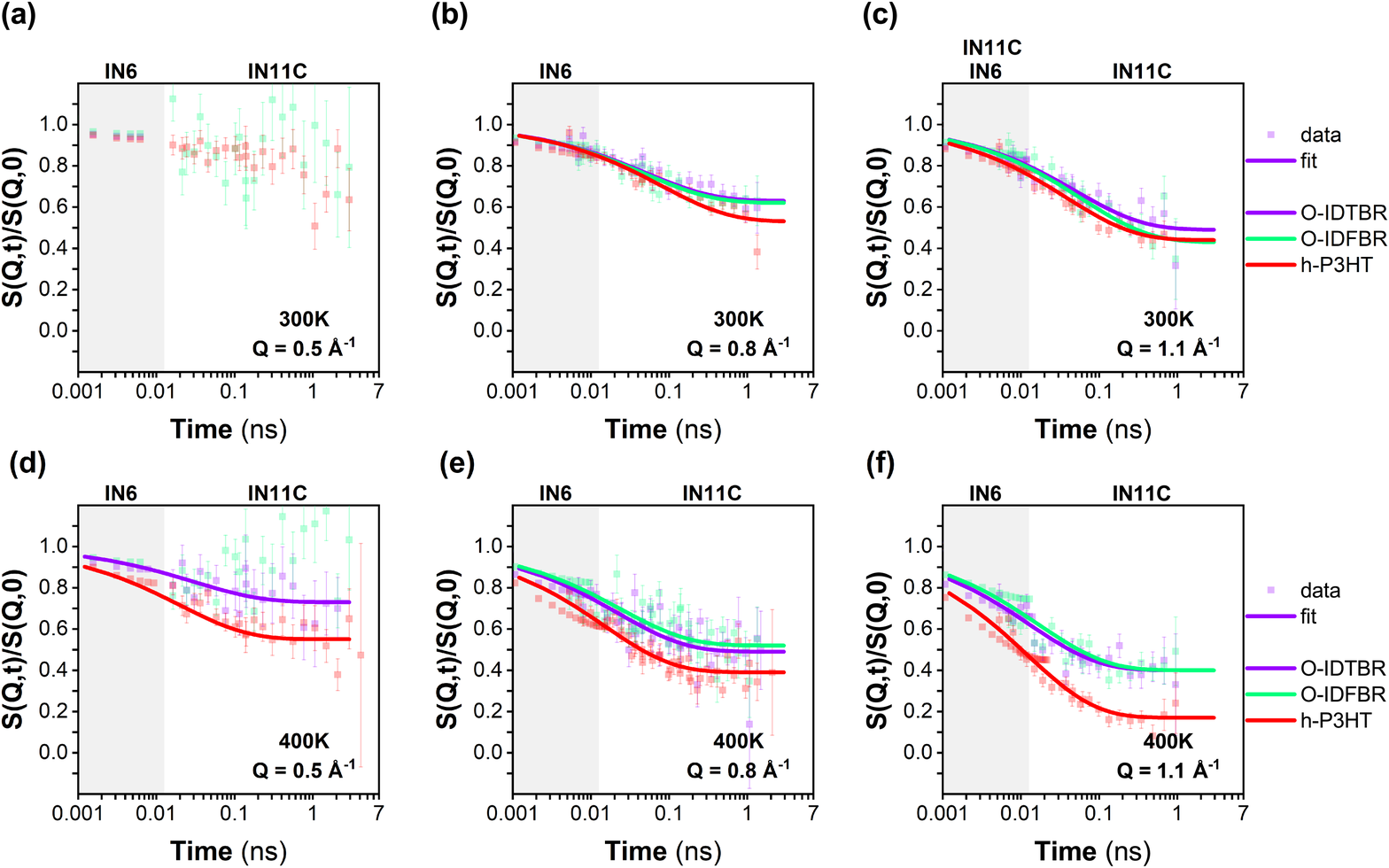}
\caption{Data and associated fits of the intermediate scattering functions of the non-fullerene acceptor molecules, O-IDTBR and O-IDFBR, and the protonated donor polymer, h-P3HT, at 300 K (a,b,c) and 400 K (d,e,f), at selected Q-values of 0.5 \AA$^{-1}$ (a,d), 0.8 \AA$^{-1}$ (b,e) and 1.1 \AA$^{-1}$ (c,f). Data were extracted from both TOF and NSE measurements, using the IN6 and IN11C spectrometers, respectively, to span a picosecond-nanosecond time window. The grey shaded area marks the accessible IN6 picosecond time window which is extended to the nanosecond region using IN11C.} 
\label{fig:fig4}
\end{figure}
Data from QENS and NSE can then be represented in the time domain as:
\begin{equation}
    \frac{S_{inc}(\mathbf{Q},t)}{S_{inc}(\mathbf{Q},0)} \approx \frac{S(\mathbf{Q},t)}{S(\mathbf{Q},0)} \approx
    \frac{\int d\omega S(\mathbf{Q}, \omega)e^{i\omega t}}{\int d\omega R(\mathbf{Q}, \omega)e^{i\omega t}}
\end{equation}
Figure \ref{fig:fig4} displays the intermediate scattering function, $\frac{S(\mathbf{Q},t)}{S(\mathbf{Q},0)}$, extracted from the inverse Fourier transform of the QENS spectra measured using the TOF spectrometer IN6 and the incoherent intermediate scattering functions measured using the NSE spectrometer IN11C. An incident neutron wavelength $\lambda_i$ = 5.12 \AA (E$_i$ = 3.12 meV) was used for IN6 measurements, offering an energy resolution at the elastic line of $\sim$ 0.07 meV, whereas on IN11C two wavelengths were used, 5.5 \AA~and 8 \AA. The IN6 time window covers up to about 50 ps, which can be extended to the nanosecond region by NSE measurements using IN11C. Data extracted from IN6 and IN11C and shown in Figure \ref{fig:fig4} are consistent and follow a coherent time-dependent trend within the covered picosecond to nanosecond window. A clear Q-dependence of the dynamics of O-IDTBR and O-IDFBR is observed. Expectedly, the relaxation accelerates as temperature increases. The relaxational dynamics of O-IDTBR and O-IDFBR on a longer timescale are similar for all Qs measured at both 300K and 400K. The observed dynamics are also closely similar to the dynamics of the polymer h-P3HT at 300K. However the situation is clearly different at 400K, where the dynamics of h-P3HT are markedly faster than the dynamics of the two NFAs for all the Qs, within the probed picosecond-nanosecond time scale matching likely motions of the side chains. If we assume that the glass transitions of the NFAs can be extrapolated from the cold crystallisation observed on differential scanning calorimetry measurements~\cite{Rezasoltani2020}, then the glass transition temperatures are $\sim$ 375K and 400K for O-IDTBR and O-IDFBR, respectively, while the glass transition temperature of P3HT is lower, around room temperature~\cite{Xie2017,Gao2022}. Therefore, the observed difference in the relaxational side chains dynamics of h-P3HT and NFAs can be assigned to the difference of their respective glass transition temperatures. Note that the backbones dynamics, slower to be captured by the present time-resolved neutron spectroscopy measurements, and the side chains dynamics, observed here, should correspond to different glass transition temperature ranges~\cite{Qian2019}; higher for the former case and lower for the latter case, respectively~\bibnote{An accurate determination of glass transition temperatures, T$_g$, is a key factor in the study of the thermal-based properties of conjugated polymers. Integrating a machine learning strategy within the experimental framework to measure T$_g$ could help achieving efficiently this goal~\cite{Alesadi2022}.}. \\
\begin{figure}
\includegraphics[width=1.1\textwidth]{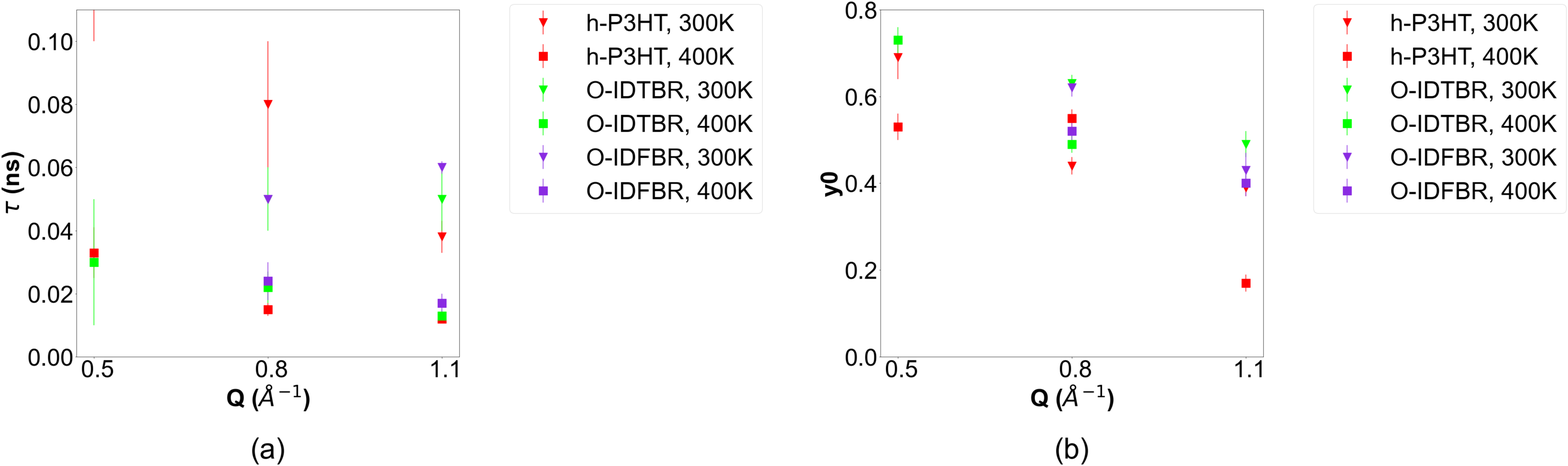}
\caption{Representation of the main fitting parameters, (a) the relaxation time $\tau$ and (b) the plateau $y_0$, as a function of Q (0.5, 0.8 and 1.1 \AA$^{-1}$) and temperature (300 and 400 K). These parameters are used in equation~\ref{stretcheq} to fit the intermediate scattering functions (Figure \ref{fig:fig4}) of the NFAs, O-IDTBR and O-IDFBR, and the protonated donor polymer h-P3HT.}
\label{fig:fig5}
\end{figure}
\\We further quantify the relaxation of the NFAs and compare with the behavior of h-P3HT by fitting $S(\mathbf{Q},t,T)$ with a stretched exponential:
\begin{equation}
\frac{S(\mathbf{Q},t,T)}{S(\mathbf{Q},0,T)} = y_{0}(\mathbf{Q,T}) + (1-y_{0}(\mathbf{Q,T})) e^{-(\frac{t}{\tau})^{\beta=\frac{1}{2}}}
\label{stretcheq}
\end{equation}
Considering a value of 0.5 for the exponent $\beta$ is reasonable for polymeric systems~\cite{Richter2005,Guilbert2019}. Figure \ref{fig:fig5} illustrates graphically the fitting parameters in equation~\ref{stretcheq}, the relaxation time $\tau$ and the plateau $y_0$. The latter is linked with the ratio of mobile/immobile scatterers as well as with the elastic incoherent structure factor (EISF). Presently, $y_0$ is observed to be smaller for h-P3HT compared to the NFAs. The relaxation time $\tau$ extracted for h-P3HT at 400K is found to be shorter than for the NFAs. In this context, the shorter relaxation time indicates further that dynamics of the side chains of h-P3HT are faster and so those of O-IDTBR and O-IDFBR are stiffer. 
\subsection{Structural dynamics of the ternary blend h-P3HT:O-IDTBR:O-IDFBR}
\begin{figure}
\includegraphics[width=0.6\textwidth]{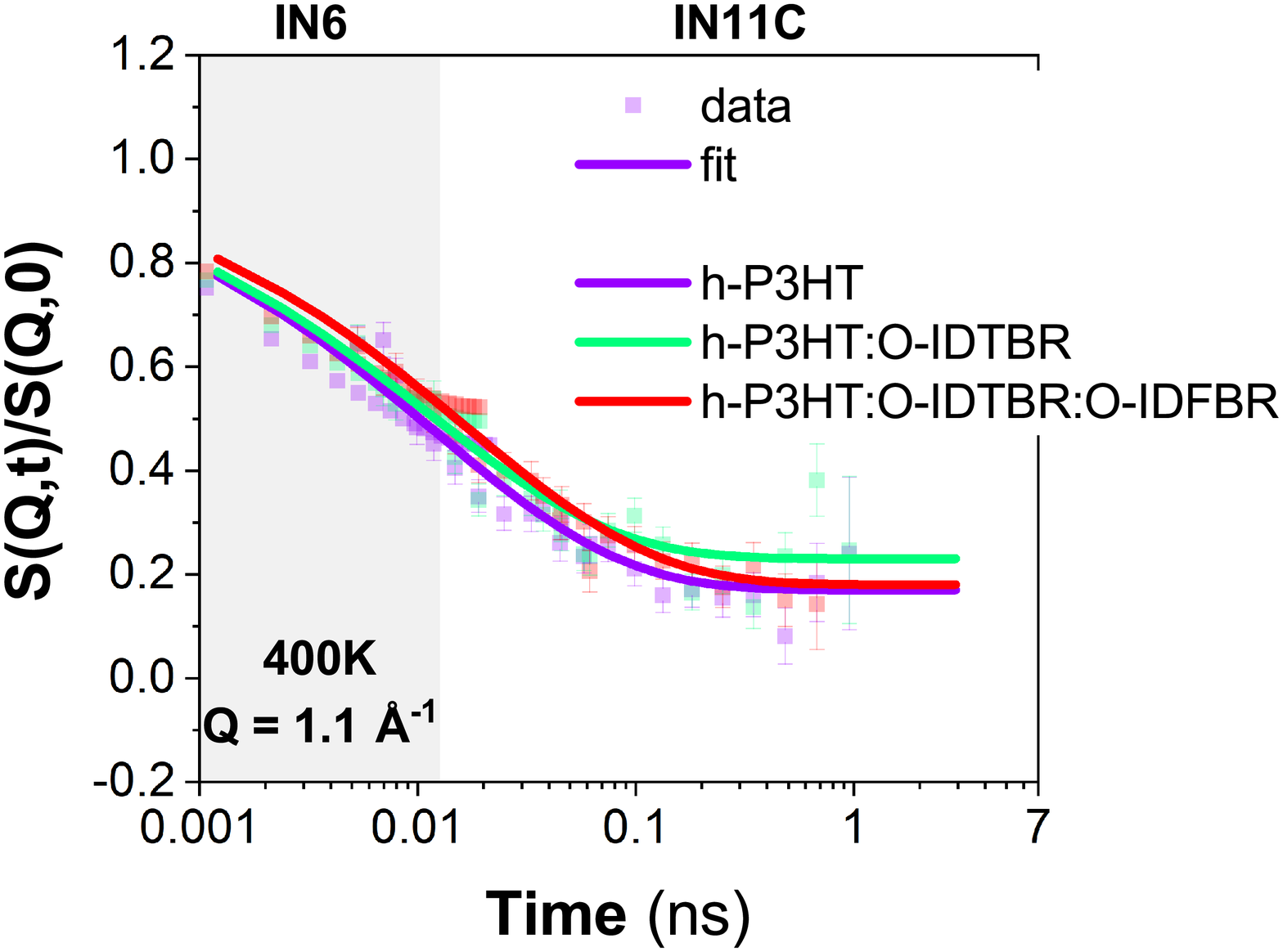}
\caption{Data and associated fits of the intermediate scattering functions at 400 K and selected Q=1.1 \AA$^{-1}$ of the protonated donor polymer h-P3HT, the binary blend h-P3HT:O-IDTBR and the ternary blend h-P3HT:O-IDTBR:O-IDFBR. Data were extracted from both TOF and NSE measurements, using the IN6 and IN11C spectrometers, respectively, to span a picosecond-nanosecond time window. The grey shaded area marks the accessible IN6 picosecond time window which is extended to the nanosecond region using IN11C.} 
\label{fig:fig6}
\end{figure}

Figure \ref{fig:fig6} displays the intermediate scattering function of the neat protonated polymer h-P3HT, the binary blend h-P3HT:O-IDTBR and the ternary blend h-P3HT:O-IDTBR:O-IDFBR at 400K for the specific Q = 1.1\AA$^{-1}$. Data were collected using both IN6 and IN11C allowing to cover an extended time window reaching the nanosecond region. No strong differences in terms of the polymer relaxation are observed upon blending with O-IDTBR and O-IDTBR:O-IDFBR despite the observed difference in dynamics between neat h-P3HT and neat NFAs reported above.\\
\begin{figure}
\includegraphics[width=\textwidth]{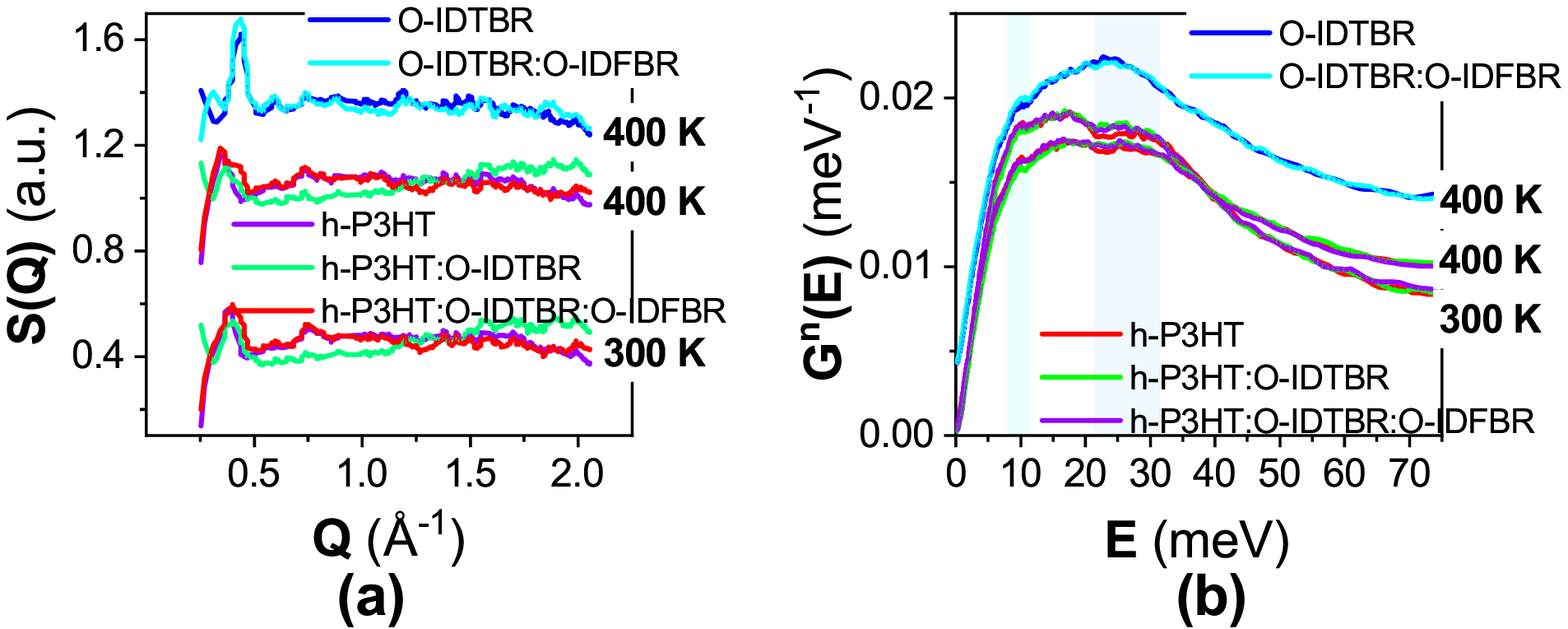}
\caption{Diffraction patterns (a) and generalized density of states (b) at 300 and 400 K of the protonated donor polymer h-P3HT, the binary donor:acceptor blend h-P3HT:O-IDTBR and the ternary donor:acceptor:acceptor blend h-P3HT:O-IDTBR:O-IDFBR, from neutron TOF-based measurements using the IN6 spectrometer, operating at an incident wavelength of 5.12~\AA ~(E$_i$=3.12 meV). Diffraction patterns and generalized density of states of the non-fullerene acceptor molecule O-IDTBR, the acceptor:acceptor blend O-IDTBR:O-IDFBR at 400K are shown as references. The light blue-shaded areas in GDOS (b) marks vibrational bands exhibiting noticeable changes upon blending with the polymer. Spectra are vertically shifted for clarity.}
\label{fig:fig7}
\end{figure}
\\Some small structural changes are observed on both neutron diffractograms (Figure \ref{fig:fig7} (a)) and generalized density of states (GDOS) (Figure \ref{fig:fig7} (b)). We previously assigned the strong diffraction peak of h-P3HT at low Q ($\sim$ 0.4 \AA$^{-1}$ at 300K) to the lamellar stacking of the polymer. The peak shifts to lower Qs ($\sim$ 0.35 \AA$^{-1}$) at 400K as can be expected from thermal expansion of the unit cell. Upon blending with O-IDTBR, the peak shifts towards higher Q-values while upon blending with O-IDTBR:O-IDFBR, the peak broadens at higher Qs at 300K and a shoulder appears at 400K. The shift and the broadening at higher Qs upon blending can be assigned to the diffraction peak of O-IDTBR at $\sim$ 0.45 \AA$^{-1}$. The differences between the binary and the ternary blends can be assigned to the peak at $\sim$ 0.3 \AA$^{-1}$ present in O-IDTBR:O-IDFBR, which can be assigned to O-IDFBR. Furthermore, differences between the GDOS of h-P3HT and the blends can be seen around 10 meV and in the $\sim$ 20-30 meV region. The latter can be attributed to the strong bend of O-IDTBR. The vibrational feature at 10 meV is stronger for h-P3HT than for O-IDTBR leading to the small changes observed upon blending. This feature is slightly stronger in the O-IDTBR:O-IDFBR blend than in the neat O-IDTBR, explaining the smaller difference in GDOS at 10 meV between h-P3HT and the ternary blend than between the neat h-P3HT and the binary blend.\\
To summarise, (i) no differences in h-P3HT dynamics are observed upon blending with O-IDTBR or O-IDTBR:O-IDFBR, (ii) the diffractograms of the binary and the ternary blends exhibit peaks that can be assigned to all three materials, and (iii) the GDOS of the binary and ternary blends are a neutron weighted average of all three phases, thus, suggesting a limited miscibility in the amorphous P3HT-rich phase.\\
To ensure that the signal is dominated by h-P3HT, we limited the content of acceptors to 25 wt\%. Therefore, the content of O-IDFBR is limited to 7.5 wt\% in the ternary blend. O-IDFBR has been shown to be more miscible than O-IDTBR and thus, we further study the binary h-P3HT:O-IDFBR and compare it with the binary h-P3HT:O-IDTBR. No significant differences are observed between h-P3HT:O-IDTBR and h-P3HT:O-IDFBR in terms of dynamics nor in terms of structure (Figures S1-4, supplementary information). With respect to the presently studied \AA{}ngstr\"om - nanometer length-scale, the nano-phase segregation, not covered here, can be reasonably assumed to occur to a larger extent with O-IDTBR on a 10s of nanometers length-scale~\cite{Rezasoltani2020,Ghasemi2019,Gao2022}. In this context, structure-oriented techniques to probe domain size and purity and phase volume fraction such as small angle neutron scattering (SANS) and resonant soft X-ray scattering
(RSoXS), which are beyond the scope of the present study focused on dynamics, allow to probe a length-scale range of $\sim$ 1 nm to $\sim$ 100 nm~\cite{Chaney2022}. Unlike with the fullerene case PCBM, no clear impact of blending with NFAs is observed on the structure and dynamics of h-P3HT. We suggest that this can be due to the differences in shape between PCBM and the studied NFAs, the former being  spherical and the latter is flat, or the differences in side chains, i.e methyl butyrate vs octyl in PCBM and the present NFAs, respectively.
\subsection{Structural dynamics of the ternary blend d-P3HT:O-IDTBR:O-IDFBR}
\begin{figure}[H]
\includegraphics[width=\textwidth]{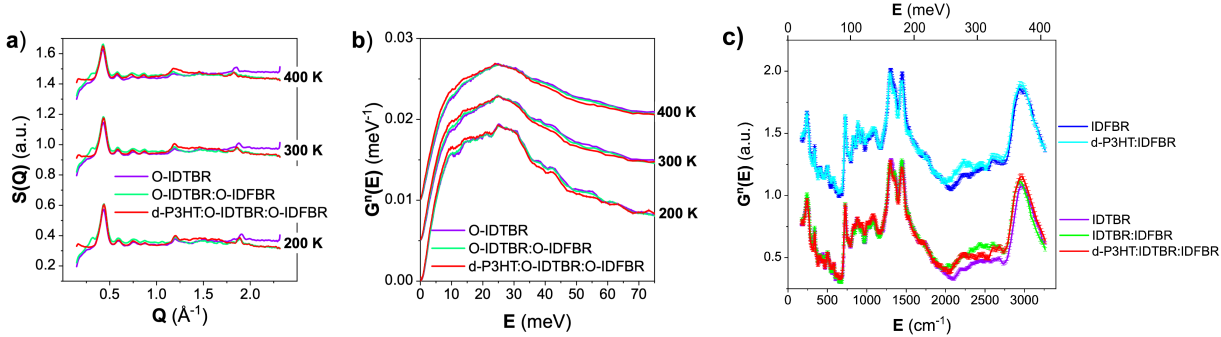}
\caption{Diffraction patterns (a) and generalized density of states (b) at 200, 300 and 400 K of the non-fullerene acceptor molecule O-IDTBR, the acceptor:acceptor blend O-IDTBR:O-IDFBR, and the ternary blend d-P3HT:O-IDTBR:O-IDFBR, with deuterated polymer d-P3HT, from neutron TOF-based measurements using the IN5 spectrometer, with an incident wavelength of 5~\AA ~(E$_i$=3.27 meV). The full-range vibrational spectra at 10 K, using the IN1-Lagrange spectrometer, of the same systems as in (a) and (b), are shown in (c), where additionally spectra of O-IDFBR and d-P3HT:O-IDFBR are also plotted as a reference. Spectra are vertically shifted for clarity.}
\label{fig:fig8}
\end{figure}
\begin{figure}[H]
\includegraphics[width=\textwidth]{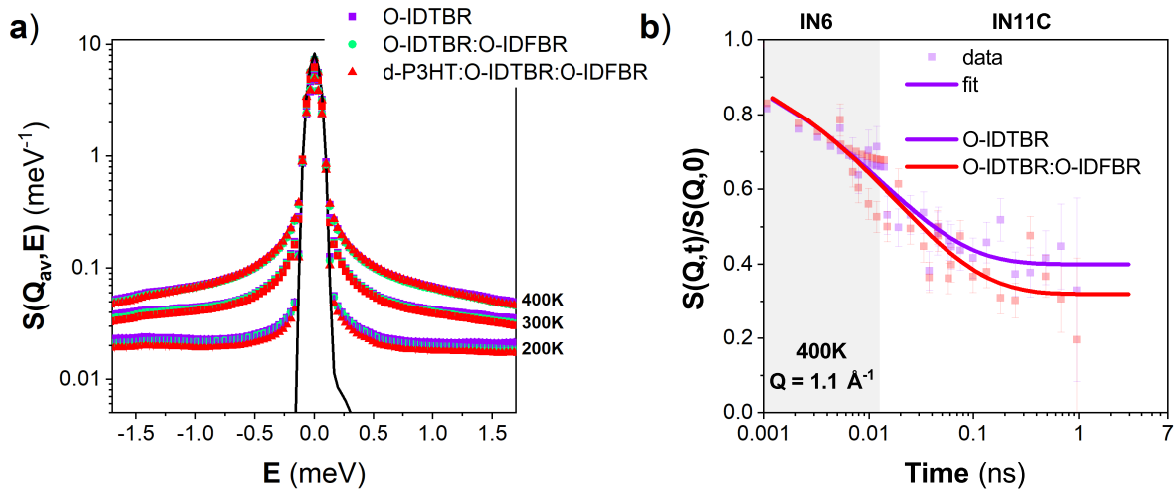}
\caption{(a) The Q-averaged TOF-based QENS spectra of O-IDTBR, O-IDTBR:O-IDFBR and O-IDTBR:d-P3HT:O-IDFBR at 200, 300 and 400 K, from TOF measurements using the IN5 spectrometer. The black line represents the instrumental resolution. (b) Data and associated fits of the intermediate scattering functions at 400 K and selected Q=1.1 \AA$^{-1}$ of O-IDTBR and O-IDTBR:O-IDFBR. Data were extracted from both TOF and NSE measurements, using the IN6 and IN11C spectrometers, respectively, to span a picosecond-nanosecond time window. The grey shaded area marks the accessible IN6 picosecond time window which is extended to the nanosecond region using IN11C.
}
\label{fig:fig9}
\end{figure}

We now characterize the dynamics of the NFAs in the blend. To do this, we took advantage of the unique capability of neutrons with regards to their contrasted interaction with isotopes of the same element and use deuterated P3HT (d-P3HT), instead of the protonated form (h-P3HT). Using d-P3HT minimizes its incoherent scattering response and therefore enabled us to better resolve the signal of the NFAs in the blends. In the diffractograms of the binary NFAs O-IDTBR:O-IDFBR (Figure \ref{fig:fig8} a), the peaks at low Q of both NFAs can be observed although the peak linked with O-IDTBR seems becoming stronger. Upon blending with d-P3HT, the peak linked with O-IDFBR at low Q disappears and broad low intensity peaks can be observed in the Q-range 1.25 - 1.75 \AA$^{-1}$, which can be assigned to d-P3HT~\cite{Guilbert2019}. From the IN5 measurements, there is no noticeable temperature effect on the anti-Stokes GDOS spectra in the 200 - 400 K range (Figure \ref{fig:fig8} b). The GDOS spectra of the neat (O-IDTBR), binary (O-IDTBR:O-IDFBR) and ternary (d-P3HT:O-IDTBR:O-IDFBR) blend resemble each other. However, some clear differences between the Stokes GDOS spectra of O-IDTBR and O-IDFBR can be observed at 10 K (Figure \ref{fig:fig8} c) in the energy range 93 - 155 meV (750 - 1250 cm$^{-1}$). This is also reflected in their binary form. Moreover, we observe a splitting of the vibrational feature around 161 meV (1300 cm$^{-1}$) upon blending O-IDTBR with O-IDFBR. Adding d-P3HT to the NFAs blend induced further changes in the GDOS as it can be observed around 19 meV ($\sim$ 153 cm$^{-1}$) in Figure \ref{fig:fig8} (b), and within 155 - 186 meV (1250 - 1500 cm$^{-1}$) in Figure \ref{fig:fig8} (c). These changes are similar to the changes observed upon blending d-P3HT with either O-IDTBR or O-IDFBR (Figure S6, supplementary information). It is thus reasonable, given the intermolecular interaction signature reflected in the vibrational spectra, to suggest that O-IDTBR forms an organic molecular alloy when mixed with O-IDFBR~\cite{Huang2019}. However, the interaction between the NFAs and P3HT is stronger, especially with O-IDFBR as evidenced by the disappearance of the Bragg peak at low Q (Figure \ref{fig:fig8} (a) and Figure S5, supplementary information, for a complementary purpose). Thus, the NFAs alloy aspect does not persist in the ternary blend with the polymer. On the relaxational dynamics side, no significant differences are observed upon blending O-IDTBR with O-IDFBR or in the ternary blend d-P3HT:O-IDTBR:O-IDFBR (Figure \ref{fig:fig9} and Figure S7, supplementary information, for a complementary purpose).\\
\\
\section{Conclusions}
In this work, we focused on probing microscopically the structural and dynamical behavior of specific binary and ternary blends, namely polymer:NFA and polymer:NFA1:NFA2, respectively. For the sake of reference and comparison, the study also included the case of a binary blend NFA1:NFA2 and its neat constituents. Neutron spectroscopy is getting recognized as increasingly valuable established tool to study dynamics of these systems at the nanometer length scale, with a suitable time scale coverage relevant to the associated molecular optoelectronic processes of the active layer. In this context, local and relaxational dynamics were investigated combining synergistically QENS and NSE. The former allowed covering the picosecond time domain, which was then extended to the nanosecond region using the latter. QENS and NSE measurements were complemented by INS to probe vibrational dynamics, within the femtosecond range. As a polymer we considered the regio-regular P3HT, under both its protonated and deuterated forms. The selected NFA molecules were O-IDTBR and O-IDFBR, under their protonated form. Deuteration allowed to tune the P3HT signal within the blend thanks to the specific isotopic sensitivity of neutrons. \\
Our findings can be summarized as follows: (i) our results confirm the difference in miscibility behavior between O-IDTBR and O-IDFBR reported previously~\cite{Rezasoltani2020}. Indeed, unlike with O-IDTBR, we observe the disappearance of the low Q-region Bragg peaks assigned to O-IDFBR upon mixing with d-P3HT, which supports that O-IDFBR is more miscible with P3HT that O-IDTBR, (ii) our neutron measurements points towards an alloying character of the NFAs blend O-IDTBR:O-IDFBR, which is not observed in the ternary blend P3HT:O-IDTBR:O-IDFBR, and (iii) interestingly, unlike what we previously observed from our studies of the fullerene-based blend P3HT:PCBM~\cite{Guilbert2016,Guilbert2017}, no differences in structural dynamics of both P3HT or NFAs are observed here upon blending within the experimentally accessible picosecond - nanosecond range, and up to 400K. In our previous work, distinct dynamical behaviors of P3HT and PCBM were observed upon blending, highlighting a vitrification/frustration of P3HT and the plasticization of PCBM, in the P3HT:PCBM blend, and which our analysis attributed to the wrapping of P3HT side chains around PCBM. Alike dynamical behaviors, involving vitrification or frustration and plasticization of one or the other component, are not presently observed in the studied NFA-based blends. We suggest that this can be assigned to the similarity in terms of chemical structures of the two NFAs O-IDTBR and O-IDFBR, and the polymer P3HT in comparison with the PCBM-based case. \\
To conclude, three important aspects, correlated with the efficiency and molecular processes of OSC active layers, have emerged from our neutron spectroscopy measurements. These are miscibility, molecular alloying and vitrification. Their presence or absence in the observations, following different blending configurations of the donor and acceptor constituents, are reflected in the probed structural dynamics at the microscopic level.\\
The change of structural dynamical behaviour between the presently studied NFAs with P3HT and the previously reported fullerene-based case, PCBM with P3HT, is striking. This opens the questions: (i) what is the role of this change in structural dynamical behavior in the improvement of the solar PCE and (ii) how much this structural dynamical behaviour is dictated by the changes in chemical structure? A dedicated thorough computational investigation would, in principle, help providing some answers by gaining deeper insights into these multifaceted aspects. As a perspective, our present studies provide important information about the relationship between chemical structure and structural dynamics and can be used to validate such a computational investigation as proposed previously~\cite{Guilbert2015,Guilbert2017,Guilbert2019}.
\section{Experimental}
\subsection*{Materials}
\begin{table*}
\small
\caption{Molecular weight (Mw), polydispersity (PDI) as measured by gel permeation chromatography and regio-regularity (RR) as measured by NMR for the protonated (h) and deuterated (d) forms of the regio-regular polymer P3HT used in this work.}
\label{tbl:tbl_1}
\begin{tabular*}{\textwidth}{@{\extracolsep{\fill}}lllllll}
\hline
 & Mw (g.mol$^{-1}$) & PDI & RR$^a$ \\
\hline
h-P3HT & 22700 & 1.19 &  90\% \\
d-P3HT & 38000 & 1.46 &  90\% \\
\hline
\end{tabular*}
\subcaption*{$^a$The RR value is assumed to also be 90\% for d-P3HT since it is synthesized by the same method as h-P3HT, but it cannot be measured as no hydrogens (replaced by deuteriums) means no 1H NMR signals.}
\end{table*}
Details of the synthesis steps of regio-regular protonated and deuterated P3HT, h-P3HT and d-P3HT, respectively, can be found in reference~\cite{Guilbert2019}. The molecular weights, polydispersities, and regio-regularities of the P3HT batches synthesized for this study are summarized in Table~\ref{tbl:tbl_1}. O-IDTBR and O-IDFBR were obtained by 1-Material Inc.
\subsection*{Neutron scattering}
\paragraph{Sample preparation}
Both the neat constituents and the blends were measured. To prepare the blends, the as-received materials were dissolved in chloroform (30 mg/mL) and drop-cast on a glass slide. The drop-cast films were then scratched from the glass substrates and stacked in aluminum foil. Each measured sample was about 400 mg. A small sample thickness of 0.2 mm was used for the neutron, spectroscopy measurements, as it is relevant to the minimization of effects like multiple scattering and absorption.
\paragraph{Quasielastic Neutron Scattering}
The temperature-dependent QENS measurements were performed at the Institut Laue-Langevin (Grenoble, France) using the direct geometry cold neutron TOF spectrometers IN5 and IN6, in a coherent way. Both IN5 and IN6 were set to operate under similar experimental conditions, with IN5 offering a better flexibility in terms of wavelength coverage not accessible on IN6. The use of IN6 was due to beamtime allocation reason to complete the initially planned measurements.  The picosecond TOF-based QENS measurements were extended to probe directly the time domain, in the nanosecond region, using the neutron spin-echo spectrometer IN11, with the IN11C 30$\degree$ detector option operating at two incident wavelengths of 5.5 and 8 \AA. On IN6, an incident wavelength of 5.12 \AA \ was used, whereas on IN5 two wavelengths of 5 and 8 \AA \ were considered. Data were collected up to 400K, hence covering the glass transitions of the studied materials, and down to 10 K for instrumental resolution purpose.
\paragraph{Inelastic Neutron Scattering} On IN5 and IN6, in addition to the QENS component, the temperature-dependent TOF measurements provide concomitantly the INS data from which we extracted the GDOS spectra in the up-scattering, neutron energy-gain mode up to $\sim$ 80 meV, and within the temperature ranges 200 – 400 K on IN5, and 150 – 400 K on IN6. The full range GDOS spectra were collected at 10 K using the hot-neutron, inverted geometry spectrometer IN1-Lagrange, in the down-scattering, neutron energy-less mode, within an accessible energy transfer range of $\sim$ 22 - 434 meV ($\sim$ 180 - 3500 cm$^{-1}$). In addition to getting the full range vibrational spectra covering both the intramolecular and intermolecular interactions, the low-temperature spectra are much less affected by the temperature-induced Debye-Waller effect leading to the broadening of the peaks as temperature increases.  
\paragraph{Neutron Diffraction} Although this is not the primary goal of neutron spectroscopy, neither intended for a quantitative purpose but rather to make a qualitative observation, diffraction patterns were also extracted from IN5 and IN6 measurements in terms the energy-averaged quasielastic component of the TOF signal.\\
Further technicalities and details of data reduction, treatment and analysis can be found in reference~\cite{Guilbert2019} regarding data collected using IN6, IN11C and IN1-Lagrange, and in reference~\cite{Zbiri2021CTF} regarding measurements using IN5. 
\section*{Author Contributions}
M.Z. and A.A.Y.G. conceived and developed the neutron project, proposed and carried out the neutron measurements, treated and analyzed the neutron data. P.A.G-F. and C.B.N. synthesized and characterized the P3HT polymers. P.F. assisted with the IN11C measurements. A.A.Y.G. and M.Z. wrote the manuscript with contribution from the co-authors.
\begin{acknowledgement}
The ILL is acknowledged for the use of the spectrometers IN5, IN6, IN11C and IN1-Lagrange. A. A. Y. G. acknowledges EPSRC for the award of an EPSRC Postdoctoral Fellowship (EP/P00928X/1).
\end{acknowledgement}
\begin{suppinfo}
Chemical compositions and incoherent neutron cross-sections of the ternary blend h-P3HT:O-IDTBR:O-IDFBR and the related three binary blend configurations, along with their neat constituents. The intermediate scattering functions of the binary blends h-P3HT:O-IDTBR and h-P3HT:O-IDFBR using the IN6 and IN11C spectrometers. The relaxation time $\tau$ and the plateau $y_0$ used in equation~5 to fit the intermediate scattering functions of h-P3HT, h-P3HT:O-IDTBR and h-P3HT:O-IDFBR. Diffraction patterns and GDOS of h-P3HT:O-IDTBR, h-P3HT:O-IDFBR and their respective constituents. Diffraction patterns and GDOS of d-P3HT:O-IDTBR, d-P3HT:O-IDFBR and their respective constituents. QENS spectra of d-P3HT:O-IDTBR, d-P3HT:O-IDFBR, O-IDTBR and O-IDFBR.
\end{suppinfo}

\providecommand{\doi}
  {\begingroup\let\do\@makeother\dospecials
  \catcode`\{=1 \catcode`\}=2\doi@aux}


\begin{mcitethebibliography}{41}
\providecommand*\natexlab[1]{#1}
\providecommand*\mciteSetBstSublistMode[1]{}
\providecommand*\mciteSetBstMaxWidthForm[2]{}
\providecommand*\mciteBstWouldAddEndPuncttrue
  {\def\EndOfBibitem{\unskip.}}
\providecommand*\mciteBstWouldAddEndPunctfalse
  {\let\EndOfBibitem\relax}
\providecommand*\mciteSetBstMidEndSepPunct[3]{}
\providecommand*\mciteSetBstSublistLabelBeginEnd[3]{}
\providecommand*\EndOfBibitem{}
\mciteSetBstSublistMode{f}
\mciteSetBstMaxWidthForm{subitem}{(\alph{mcitesubitemcount})}
\mciteSetBstSublistLabelBeginEnd
  {\mcitemaxwidthsubitemform\space}
  {\relax}
  {\relax}

\bibitem[Armin \latin{et~al.}(2021)Armin, Li, Sandberg, Xiao, Ding, Nelson,
  Neher, Vandewal, Shoaee, Wang, Ade, HeumÃ¼ller, Brabec, and
  Meredith]{Armin2021}
Armin,~A.; Li,~W.; Sandberg,~O.~J.; Xiao,~Z.; Ding,~L.; Nelson,~J.; Neher,~D.;
  Vandewal,~K.; Shoaee,~S.; Wang,~T.; Ade,~H.; HeumÃ¼ller,~T.; Brabec,~C.;
  Meredith,~P. A History and Perspective of Non-Fullerene Electron Acceptors
  for Organic Solar Cells. \emph{Advanced Energy Materials} \textbf{2021},
  \emph{11}, 2003570
\mciteBstWouldAddEndPuncttrue
\mciteSetBstMidEndSepPunct{\mcitedefaultmidpunct}
{\mcitedefaultendpunct}{\mcitedefaultseppunct}\relax
\EndOfBibitem
\bibitem[Naveed and Ma(2018)Naveed, and Ma]{Naveed2018}
Naveed,~H.~B.; Ma,~W. Miscibility-Driven Optimization of Nanostructures in
  Ternary Organic Solar Cells Using Non-fullerene Acceptors. \emph{Joule}
  \textbf{2018}, \emph{2}, 621--641
\mciteBstWouldAddEndPuncttrue
\mciteSetBstMidEndSepPunct{\mcitedefaultmidpunct}
{\mcitedefaultendpunct}{\mcitedefaultseppunct}\relax
\EndOfBibitem
\bibitem[Gillett \latin{et~al.}(2021)Gillett, Privitera, Dilmurat, Karki, Qian,
  Pershin, Londi, Myers, Lee, Yuan, Ko, Riede, Gao, Bazan, Rao, Nguyen,
  Beljonne, and Friend]{Gillett2021}
Gillett,~A.~J.; Privitera,~A.; Dilmurat,~R.; Karki,~A.; Qian,~D.; Pershin,~A.;
  Londi,~G.; Myers,~W.~K.; Lee,~J.; Yuan,~J.; Ko,~S.-J.; Riede,~M.~K.; Gao,~F.;
  Bazan,~G.~C.; Rao,~A.; Nguyen,~T.-Q.; Beljonne,~D.; Friend,~R.~H. The role of
  charge recombination to triplet excitons in organic solar cells.
  \emph{Nature} \textbf{2021}, \emph{597}, 666--671
\mciteBstWouldAddEndPuncttrue
\mciteSetBstMidEndSepPunct{\mcitedefaultmidpunct}
{\mcitedefaultendpunct}{\mcitedefaultseppunct}\relax
\EndOfBibitem
\bibitem[Gao \latin{et~al.}(2022)Gao, Liu, Xian, Peng, Zhou, Liu, Li, Xie,
  Zhao, Zhang, Jiao, and Ye]{Gao2022}
Gao,~M.; Liu,~Y.; Xian,~K.; Peng,~Z.; Zhou,~K.; Liu,~J.; Li,~S.; Xie,~F.;
  Zhao,~W.; Zhang,~J.; Jiao,~X.; Ye,~L. Thermally stable
  poly(3-hexylthiophene): Nonfullerene solar cells with efficiency breaking
  10\%. \emph{Aggregate} \textbf{2022}, \emph{3}, e190
\mciteBstWouldAddEndPuncttrue
\mciteSetBstMidEndSepPunct{\mcitedefaultmidpunct}
{\mcitedefaultendpunct}{\mcitedefaultseppunct}\relax
\EndOfBibitem
\bibitem[Zbiri \latin{et~al.}(2021)Zbiri, Finn, Nielsen, and
  Guilbert]{Zbiri2021}
Zbiri,~M.; Finn,~P.~A.; Nielsen,~C.~B.; Guilbert,~A. A.~Y. Quantitative
  insights into the phase behaviour and miscibility of organic photovoltaic
  active layers from the perspective of neutron spectroscopy. \emph{J. Mater.
  Chem. C} \textbf{2021}, \emph{9}, 11873--11881
\mciteBstWouldAddEndPuncttrue
\mciteSetBstMidEndSepPunct{\mcitedefaultmidpunct}
{\mcitedefaultendpunct}{\mcitedefaultseppunct}\relax
\EndOfBibitem
\bibitem[Müller \latin{et~al.}(2008)Müller, Ferenczi, Campoy-Quiles, Frost,
  Bradley, Smith, Stingelin-Stutzmann, and Nelson]{Muller2008}
Müller,~C.; Ferenczi,~T. A.~M.; Campoy-Quiles,~M.; Frost,~J.~M.; Bradley,~D.
  D.~C.; Smith,~P.; Stingelin-Stutzmann,~N.; Nelson,~J. Binary Organic
  Photovoltaic Blends: A Simple Rationale for Optimum Compositions.
  \emph{Advanced Materials} \textbf{2008}, \emph{20}, 3510--3515
\mciteBstWouldAddEndPuncttrue
\mciteSetBstMidEndSepPunct{\mcitedefaultmidpunct}
{\mcitedefaultendpunct}{\mcitedefaultseppunct}\relax
\EndOfBibitem
\bibitem[Wadsworth \latin{et~al.}(2018)Wadsworth, Hamid, Bidwell, Ashraf, Khan,
  Anjum, Cendra, Yan, Rezasoltani, Guilbert, Azzouzi, Gasparini, Bannock,
  Baran, Wu, de~Mello, Brabec, Salleo, Nelson, Laquai, and
  McCulloch]{Wadsworth2018}
Wadsworth,~A.; Hamid,~Z.; Bidwell,~M.; Ashraf,~R.~S.; Khan,~J.~I.;
  Anjum,~D.~H.; Cendra,~C.; Yan,~J.; Rezasoltani,~E.; Guilbert,~A. A.~Y.;
  Azzouzi,~M.; Gasparini,~N.; Bannock,~J.~H.; Baran,~D.; Wu,~H.;
  de~Mello,~J.~C.; Brabec,~C.~J.; Salleo,~A.; Nelson,~J.; Laquai,~F.;
  McCulloch,~I. Progress in Poly (3-Hexylthiophene) Organic Solar Cells and the
  Influence of Its Molecular Weight on Device Performance. \emph{Advanced
  Energy Materials} \textbf{2018}, \emph{8}, 1801001
\mciteBstWouldAddEndPuncttrue
\mciteSetBstMidEndSepPunct{\mcitedefaultmidpunct}
{\mcitedefaultendpunct}{\mcitedefaultseppunct}\relax
\EndOfBibitem
\bibitem[Rezasoltani \latin{et~al.}(2020)Rezasoltani, Guilbert, Yan,
  Rodríguez-Martínez, Azzouzi, Eisner, Tuladhar, Hamid, Wadsworth, McCulloch,
  Campoy-Quiles, and Nelson]{Rezasoltani2020}
Rezasoltani,~E.; Guilbert,~A.; Yan,~J.; Rodríguez-Martínez,~X.; Azzouzi,~M.;
  Eisner,~F.; Tuladhar,~S.; Hamid,~Z.; Wadsworth,~A.; McCulloch,~I.;
  Campoy-Quiles,~M.; Nelson,~J. Correlating the Phase Behavior with the Device
  Performance in Binary Poly-3-hexylthiophene: Nonfullerene Acceptor Blend
  Using Optical Probes of the Microstructure. \emph{Chemistry of Materials}
  \textbf{2020}, \emph{32}, 8294--8305
\mciteBstWouldAddEndPuncttrue
\mciteSetBstMidEndSepPunct{\mcitedefaultmidpunct}
{\mcitedefaultendpunct}{\mcitedefaultseppunct}\relax
\EndOfBibitem
\bibitem[Guilbert \latin{et~al.}(2012)Guilbert, Reynolds, Bruno, Maclachlan,
  King, Faist, Pires, MacDonald, Stingelin, Haque, and Nelson]{Guilbert2012}
Guilbert,~A.; Reynolds,~L.; Bruno,~A.; Maclachlan,~A.; King,~S.; Faist,~M.;
  Pires,~E.; MacDonald,~J.; Stingelin,~N.; Haque,~S.; Nelson,~J. Effect of
  multiple adduct fullerenes on microstructure and phase behavior of P3HT:
  Fullerene blend films for organic solar cells. \emph{ACS Nano} \textbf{2012},
  \emph{6}, 3868–3875
\mciteBstWouldAddEndPuncttrue
\mciteSetBstMidEndSepPunct{\mcitedefaultmidpunct}
{\mcitedefaultendpunct}{\mcitedefaultseppunct}\relax
\EndOfBibitem
\bibitem[Ghasemi \latin{et~al.}(2019)Ghasemi, Hu, Peng, Rech, Angunawela,
  Carpenter, Stuard, Wadsworth, McCulloch, You, and Ade]{Ghasemi2019}
Ghasemi,~M.; Hu,~H.; Peng,~Z.; Rech,~J.~J.; Angunawela,~I.; Carpenter,~J.~H.;
  Stuard,~S.~J.; Wadsworth,~A.; McCulloch,~I.; You,~W.; Ade,~H. Delineation of
  Thermodynamic and Kinetic Factors that Control Stability in Non-fullerene
  Organic Solar Cells. \emph{Joule} \textbf{2019}, \emph{3}, 1328--1348
\mciteBstWouldAddEndPuncttrue
\mciteSetBstMidEndSepPunct{\mcitedefaultmidpunct}
{\mcitedefaultendpunct}{\mcitedefaultseppunct}\relax
\EndOfBibitem
\bibitem[Saladina \latin{et~al.}(2021)Saladina, Simón~Marqués, Markina,
  Karuthedath, Wöpke, Göhler, Chen, Allain, Blanchard, Cabanetos, Andrienko,
  Laquai, Gorenflot, and Deibel]{Saladina2021}
Saladina,~M.; Simón~Marqués,~P.; Markina,~A.; Karuthedath,~S.; Wöpke,~C.;
  Göhler,~C.; Chen,~Y.; Allain,~M.; Blanchard,~P.; Cabanetos,~C.;
  Andrienko,~D.; Laquai,~F.; Gorenflot,~J.; Deibel,~C. Charge Photogeneration
  in Non-Fullerene Organic Solar Cells: Influence of Excess Energy and
  Electrostatic Interactions. \emph{Advanced Functional Materials}
  \textbf{2021}, \emph{31}, 2007479
\mciteBstWouldAddEndPuncttrue
\mciteSetBstMidEndSepPunct{\mcitedefaultmidpunct}
{\mcitedefaultendpunct}{\mcitedefaultseppunct}\relax
\EndOfBibitem
\bibitem[Ghasemi \latin{et~al.}(2021)Ghasemi, Balar, Peng, Hu, Qin, Kim, Rech,
  Bidwell, Mask, McCulloch, You, Amassian, Risko, O’Connor, and
  Ade]{Ghasemi2021}
Ghasemi,~M.; Balar,~N.; Peng,~Z.; Hu,~H.; Qin,~Y.; Kim,~T.; Rech,~J.~J.;
  Bidwell,~M.; Mask,~W.; McCulloch,~I.; You,~W.; Amassian,~A.; Risko,~C.;
  O’Connor,~B.~T.; Ade,~H. A molecular interaction–diffusion framework for
  predicting organic solar cell stability. \emph{Nature Materials}
  \textbf{2021}, \emph{20}, 525--532
\mciteBstWouldAddEndPuncttrue
\mciteSetBstMidEndSepPunct{\mcitedefaultmidpunct}
{\mcitedefaultendpunct}{\mcitedefaultseppunct}\relax
\EndOfBibitem
\bibitem[Baran \latin{et~al.}(2017)Baran, Ashraf, Hanifi, Abdelsamie,
  Gasparini, R{\"o}hr, Holliday, Wadsworth, Lockett, Neophytou, Emmott, Nelson,
  Brabec, Amassian, Salleo, Kirchartz, Durrant, and McCulloch]{Baran2017}
Baran,~D.; Ashraf,~R.~S.; Hanifi,~D.~A.; Abdelsamie,~M.; Gasparini,~N.;
  R{\"o}hr,~J.~A.; Holliday,~S.; Wadsworth,~A.; Lockett,~S.; Neophytou,~M.;
  Emmott,~C. J.~M.; Nelson,~J.; Brabec,~C.~J.; Amassian,~A.; Salleo,~A.;
  Kirchartz,~T.; Durrant,~J.~R.; McCulloch,~I. Reducing the
  efficiency-stability-cost gap of organic photovoltaics with highly efficient
  and stable small molecule acceptor ternary solar cells. \emph{Nature
  Materials} \textbf{2017}, \emph{16}, 363--369
\mciteBstWouldAddEndPuncttrue
\mciteSetBstMidEndSepPunct{\mcitedefaultmidpunct}
{\mcitedefaultendpunct}{\mcitedefaultseppunct}\relax
\EndOfBibitem
\bibitem[Chen \latin{et~al.}(2017)Chen, Liu, Zhang, Chow, Wang, Zhang, Ma, and
  Yan]{Chen2017}
Chen,~S.; Liu,~Y.; Zhang,~L.; Chow,~P. C.~Y.; Wang,~Z.; Zhang,~G.; Ma,~W.;
  Yan,~H. A Wide-Bandgap Donor Polymer for Highly Efficient Non-fullerene
  Organic Solar Cells with a Small Voltage Loss. \emph{Journal of the American
  Chemical Society} \textbf{2017}, \emph{139}, 6298--6301
\mciteBstWouldAddEndPuncttrue
\mciteSetBstMidEndSepPunct{\mcitedefaultmidpunct}
{\mcitedefaultendpunct}{\mcitedefaultseppunct}\relax
\EndOfBibitem
\bibitem[Gasparini \latin{et~al.}(2017)Gasparini, Salvador, Heumueller,
  Richter, Classen, Shrestha, Matt, Holliday, Strohm, Egelhaaf, Wadsworth,
  Baran, McCulloch, and Brabec]{Gasparini2017}
Gasparini,~N.; Salvador,~M.; Heumueller,~T.; Richter,~M.; Classen,~A.;
  Shrestha,~S.; Matt,~G.~J.; Holliday,~S.; Strohm,~S.; Egelhaaf,~H.-J.;
  Wadsworth,~A.; Baran,~D.; McCulloch,~I.; Brabec,~C.~J. Polymer:Nonfullerene
  Bulk Heterojunction Solar Cells with Exceptionally Low Recombination Rates.
  \emph{Advanced Energy Materials} \textbf{2017}, \emph{7}, 1701561
\mciteBstWouldAddEndPuncttrue
\mciteSetBstMidEndSepPunct{\mcitedefaultmidpunct}
{\mcitedefaultendpunct}{\mcitedefaultseppunct}\relax
\EndOfBibitem
\bibitem[Liang \latin{et~al.}(2018)Liang, Babics, Savikhin, Zhang, Le~Corre,
  Lopatin, Kan, Firdaus, Liu, McCulloch, Toney, and Beaujuge]{Liang2018}
Liang,~R.-Z.; Babics,~M.; Savikhin,~V.; Zhang,~W.; Le~Corre,~V.~M.;
  Lopatin,~S.; Kan,~Z.; Firdaus,~Y.; Liu,~S.; McCulloch,~I.; Toney,~M.~F.;
  Beaujuge,~P.~M. Carrier Transport and Recombination in Efficient
  “All-Small-Molecule” Solar Cells with the Nonfullerene Acceptor IDTBR.
  \emph{Advanced Energy Materials} \textbf{2018}, \emph{8}, 1800264
\mciteBstWouldAddEndPuncttrue
\mciteSetBstMidEndSepPunct{\mcitedefaultmidpunct}
{\mcitedefaultendpunct}{\mcitedefaultseppunct}\relax
\EndOfBibitem
\bibitem[Hoefler \latin{et~al.}(2018)Hoefler, Rath, Pastukhova, Pavlica,
  Scheunemann, Wilken, Kunert, Resel, Hobisch, Xiao, Bratina, and
  Trimmel]{Hoefler2018}
Hoefler,~S.~F.; Rath,~T.; Pastukhova,~N.; Pavlica,~E.; Scheunemann,~D.;
  Wilken,~S.; Kunert,~B.; Resel,~R.; Hobisch,~M.; Xiao,~S.; Bratina,~G.;
  Trimmel,~G. The effect of polymer molecular weight on the performance of
  PTB7-Th:O-IDTBR non-fullerene organic solar cells. \emph{J. Mater. Chem. A}
  \textbf{2018}, \emph{6}, 9506--9516
\mciteBstWouldAddEndPuncttrue
\mciteSetBstMidEndSepPunct{\mcitedefaultmidpunct}
{\mcitedefaultendpunct}{\mcitedefaultseppunct}\relax
\EndOfBibitem
\bibitem[Corzo \latin{et~al.}(2019)Corzo, Almasabi, Bihar, Macphee,
  Rosas-Villalva, Gasparini, Inal, and Baran]{Corzo2019}
Corzo,~D.; Almasabi,~K.; Bihar,~E.; Macphee,~S.; Rosas-Villalva,~D.;
  Gasparini,~N.; Inal,~S.; Baran,~D. Digital Inkjet Printing of High-Efficiency
  Large-Area Nonfullerene Organic Solar Cells. \emph{Advanced Materials
  Technologies} \textbf{2019}, \emph{4}, 1900040
\mciteBstWouldAddEndPuncttrue
\mciteSetBstMidEndSepPunct{\mcitedefaultmidpunct}
{\mcitedefaultendpunct}{\mcitedefaultseppunct}\relax
\EndOfBibitem
\bibitem[López-Vicente \latin{et~al.}(2021)López-Vicente, Fernández-Castro,
  Abad, Mazzolini, Andreasen, Espindola-Rodriguez, and
  Urbina]{Lopez-Vicente2021}
López-Vicente,~R.; Fernández-Castro,~M.; Abad,~J.; Mazzolini,~E.;
  Andreasen,~J.~W.; Espindola-Rodriguez,~M.; Urbina,~A. Lifetime Study of
  Organic Solar Cells with O-IDTBR as Non-Fullerene Acceptor. \emph{Frontiers
  in Energy Research} \textbf{2021}, \emph{9}, 741288
\mciteBstWouldAddEndPuncttrue
\mciteSetBstMidEndSepPunct{\mcitedefaultmidpunct}
{\mcitedefaultendpunct}{\mcitedefaultseppunct}\relax
\EndOfBibitem
\bibitem[Yan \latin{et~al.}(2021)Yan, Rezasoltani, Azzouzi, Eisner, and
  Nelson]{Yan2021}
Yan,~J.; Rezasoltani,~E.; Azzouzi,~M.; Eisner,~F.; Nelson,~J. Influence of
  static disorder of charge transfer state on voltage loss in organic
  photovoltaics. \emph{Nature Communications} \textbf{2021}, \emph{12}, 3642
\mciteBstWouldAddEndPuncttrue
\mciteSetBstMidEndSepPunct{\mcitedefaultmidpunct}
{\mcitedefaultendpunct}{\mcitedefaultseppunct}\relax
\EndOfBibitem
\bibitem[Cavaye(2019)]{Cavaye2019}
Cavaye,~H. Neutron Spectroscopy: An Under-Utilised Tool for Organic Electronics
  Research? \emph{Angewandte Chemie International Edition} \textbf{2019},
  \emph{58}, 9338--9346
\mciteBstWouldAddEndPuncttrue
\mciteSetBstMidEndSepPunct{\mcitedefaultmidpunct}
{\mcitedefaultendpunct}{\mcitedefaultseppunct}\relax
\EndOfBibitem
\bibitem[Guilbert \latin{et~al.}(2016)Guilbert, Zbiri, Jenart, Nielsen, and
  Nelson]{Guilbert2016}
Guilbert,~A.; Zbiri,~M.; Jenart,~M.; Nielsen,~C.; Nelson,~J. New Insights into
  the Molecular Dynamics of P3HT:PCBM Bulk Heterojunction: A Time-of-Flight
  Quasi-Elastic Neutron Scattering Study. \emph{Journal of Physical Chemistry
  Letters} \textbf{2016}, \emph{7}, 2252–2257
\mciteBstWouldAddEndPuncttrue
\mciteSetBstMidEndSepPunct{\mcitedefaultmidpunct}
{\mcitedefaultendpunct}{\mcitedefaultseppunct}\relax
\EndOfBibitem
\bibitem[Guilbert \latin{et~al.}(2017)Guilbert, Zbiri, Dunbar, and
  Nelson]{Guilbert2017}
Guilbert,~A.; Zbiri,~M.; Dunbar,~A.; Nelson,~J. Quantitative Analysis of the
  Molecular Dynamics of P3HT:PCBM Bulk Heterojunction. \emph{Journal of
  Physical Chemistry B} \textbf{2017}, \emph{121}, 9073–9080
\mciteBstWouldAddEndPuncttrue
\mciteSetBstMidEndSepPunct{\mcitedefaultmidpunct}
{\mcitedefaultendpunct}{\mcitedefaultseppunct}\relax
\EndOfBibitem
\bibitem[Guilbert \latin{et~al.}(2015)Guilbert, Urbina, Abad, Díaz-Paniagua,
  Batallán, Seydel, Zbiri, García-Sakai, and Nelson]{Guilbert2015}
Guilbert,~A. A.~Y.; Urbina,~A.; Abad,~J.; Díaz-Paniagua,~C.; Batallán,~F.;
  Seydel,~T.; Zbiri,~M.; García-Sakai,~V.; Nelson,~J. Temperature-Dependent
  Dynamics of Polyalkylthiophene Conjugated Polymers: A Combined Neutron
  Scattering and Simulation Study. \emph{Chemistry of Materials} \textbf{2015},
  \emph{27}, 7652--7661
\mciteBstWouldAddEndPuncttrue
\mciteSetBstMidEndSepPunct{\mcitedefaultmidpunct}
{\mcitedefaultendpunct}{\mcitedefaultseppunct}\relax
\EndOfBibitem
\bibitem[Guilbert \latin{et~al.}(2019)Guilbert, Zbiri, Finn, Jenart, Fouquet,
  Cristiglio, Frick, Nelson, and Nielsen]{Guilbert2019}
Guilbert,~A.~A.; Zbiri,~M.; Finn,~P.~A.; Jenart,~M.; Fouquet,~P.;
  Cristiglio,~V.; Frick,~B.; Nelson,~J.; Nielsen,~C.~B. Mapping Microstructural
  Dynamics up to the Nanosecond of the Conjugated Polymer P3HT in the Solid
  State. \emph{Chemistry of Materials} \textbf{2019}, \emph{31}, 9635--9651
\mciteBstWouldAddEndPuncttrue
\mciteSetBstMidEndSepPunct{\mcitedefaultmidpunct}
{\mcitedefaultendpunct}{\mcitedefaultseppunct}\relax
\EndOfBibitem
\bibitem[Guilbert \latin{et~al.}(2021)Guilbert, Parr, Kreouzis, Woods, Sprick,
  Abrahams, Nielsen, and Zbiri]{Guilbert2021PTTPFS}
Guilbert,~A. A.~Y.; Parr,~Z.~S.; Kreouzis,~T.; Woods,~D.~J.; Sprick,~R.~S.;
  Abrahams,~I.; Nielsen,~C.~B.; Zbiri,~M. Effect of substituting non-polar
  chains with polar chains on the structural dynamics of small organic molecule
  and polymer semiconductors. \emph{Phys. Chem. Chem. Phys.} \textbf{2021},
  \emph{23}, 7462--7471
\mciteBstWouldAddEndPuncttrue
\mciteSetBstMidEndSepPunct{\mcitedefaultmidpunct}
{\mcitedefaultendpunct}{\mcitedefaultseppunct}\relax
\EndOfBibitem
\bibitem[Stoeckel \latin{et~al.}(2021)Stoeckel, Olivier, Gobbi, Dudenko,
  Lemaur, Zbiri, Guilbert, D'Avino, Liscio, Migliori, Ortolani, Demitri, Jin,
  Jeong, Liscio, Nardi, Pasquali, Razzari, Beljonne, SamorÃ¬, and
  Orgiu]{Stoeckel2021}
Stoeckel,~M.-A.; Olivier,~Y.; Gobbi,~M.; Dudenko,~D.; Lemaur,~V.; Zbiri,~M.;
  Guilbert,~A. A.~Y.; D'Avino,~G.; Liscio,~F.; Migliori,~A.; Ortolani,~L.;
  Demitri,~N.; Jin,~X.; Jeong,~Y.-G.; Liscio,~A.; Nardi,~M.-V.; Pasquali,~L.;
  Razzari,~L.; Beljonne,~D.; SamorÃ¬,~P.; Orgiu,~E. Analysis of External and
  Internal Disorder to Understand Band-Like Transport in n-Type Organic
  Semiconductors. \emph{Advanced Materials} \textbf{2021}, \emph{33}, 2007870
\mciteBstWouldAddEndPuncttrue
\mciteSetBstMidEndSepPunct{\mcitedefaultmidpunct}
{\mcitedefaultendpunct}{\mcitedefaultseppunct}\relax
\EndOfBibitem
\bibitem[Guilbert \latin{et~al.}(2021)Guilbert, Bai, Aitchison, Sprick, and
  Zbiri]{Guilbert2021CMP}
Guilbert,~A. A.~Y.; Bai,~Y.; Aitchison,~C.~M.; Sprick,~R.~S.; Zbiri,~M. Impact
  of Chemical Structure on the Dynamics of Mass Transfer of Water in Conjugated
  Microporous Polymers: A Neutron Spectroscopy Study. \emph{ACS Applied Polymer
  Materials} \textbf{2021}, \emph{3}, 765--776
\mciteBstWouldAddEndPuncttrue
\mciteSetBstMidEndSepPunct{\mcitedefaultmidpunct}
{\mcitedefaultendpunct}{\mcitedefaultseppunct}\relax
\EndOfBibitem
\bibitem[Sprick \latin{et~al.}(2019)Sprick, Bai, Guilbert, Zbiri, Aitchison,
  Wilbraham, Yan, Woods, Zwijnenburg, and Cooper]{Sprick2019}
Sprick,~R.~S.; Bai,~Y.; Guilbert,~A. A.~Y.; Zbiri,~M.; Aitchison,~C.~M.;
  Wilbraham,~L.; Yan,~Y.; Woods,~D.~J.; Zwijnenburg,~M.~A.; Cooper,~A.~I.
  Photocatalytic Hydrogen Evolution from Water Using Fluorene and
  Dibenzothiophene Sulfone-Conjugated Microporous and Linear Polymers.
  \emph{Chemistry of Materials} \textbf{2019}, \emph{31}, 305--313
\mciteBstWouldAddEndPuncttrue
\mciteSetBstMidEndSepPunct{\mcitedefaultmidpunct}
{\mcitedefaultendpunct}{\mcitedefaultseppunct}\relax
\EndOfBibitem
\bibitem[Zbiri \latin{et~al.}(2021)Zbiri, Aitchison, Sprick, Cooper, and
  Guilbert]{Zbiri2021CTF}
Zbiri,~M.; Aitchison,~C.~M.; Sprick,~R.~S.; Cooper,~A.~I.; Guilbert,~A. A.~Y.
  Probing Dynamics of Water Mass Transfer in Organic Porous Photocatalyst
  Water-Splitting Materials by Neutron Spectroscopy. \emph{Chemistry of
  Materials} \textbf{2021}, \emph{33}, 1363--1372
\mciteBstWouldAddEndPuncttrue
\mciteSetBstMidEndSepPunct{\mcitedefaultmidpunct}
{\mcitedefaultendpunct}{\mcitedefaultseppunct}\relax
\EndOfBibitem
\bibitem[Not()]{Note-1}
This is in contrast with optical spectroscopy, which is subjected to
  photoluminescence when dealing with conjugated organic materials. The light
  probe interacts with the optical and electronic processes of these systems.
  In this context, the absorption range of the sample and/or the measured
  signal contaminated with a contribution from subsequent light emission of the
  material can limit or make it difficult to use optical spectroscopy to study
  dynamics of organic semiconductors.\relax
\mciteBstWouldAddEndPunctfalse
\mciteSetBstMidEndSepPunct{\mcitedefaultmidpunct}
{}{\mcitedefaultseppunct}\relax
\EndOfBibitem
\bibitem[Bristow \latin{et~al.}(2019)Bristow, Thorley, White, Wadsworth,
  Babics, Hamid, Zhang, Paterson, Kosco, Panidi, Anthopoulos, and
  McCulloch]{Bristow2019}
Bristow,~H.; Thorley,~K.~J.; White,~A.~J.; Wadsworth,~A.; Babics,~M.;
  Hamid,~Z.; Zhang,~W.; Paterson,~A.~F.; Kosco,~J.; Panidi,~J.;
  Anthopoulos,~T.~D.; McCulloch,~I. Impact of Nonfullerene Acceptor Side Chain
  Variation on Transistor Mobility. \emph{Advanced Electronic Materials}
  \textbf{2019}, \emph{5}, 1900344
\mciteBstWouldAddEndPuncttrue
\mciteSetBstMidEndSepPunct{\mcitedefaultmidpunct}
{\mcitedefaultendpunct}{\mcitedefaultseppunct}\relax
\EndOfBibitem
\bibitem[Halaby \latin{et~al.}(2021)Halaby, Martynowycz, Zhu, Tretiak,
  Zhugayevych, Gonen, and Seifrid]{Halaby2021}
Halaby,~S.; Martynowycz,~M.~W.; Zhu,~Z.; Tretiak,~S.; Zhugayevych,~A.;
  Gonen,~T.; Seifrid,~M. Microcrystal Electron Diffraction for Molecular Design
  of Functional Non-Fullerene Acceptor Structures. \emph{Chemistry of
  Materials} \textbf{2021}, \emph{33}, 966--977
\mciteBstWouldAddEndPuncttrue
\mciteSetBstMidEndSepPunct{\mcitedefaultmidpunct}
{\mcitedefaultendpunct}{\mcitedefaultseppunct}\relax
\EndOfBibitem
\bibitem[Xie \latin{et~al.}(2017)Xie, Lee, Aplan, Caggiano, Müller, Colby, and
  Gomez]{Xie2017}
Xie,~R.; Lee,~Y.; Aplan,~M.~P.; Caggiano,~N.~J.; Müller,~C.; Colby,~R.~H.;
  Gomez,~E.~D. Glass Transition Temperature of Conjugated Polymers by
  Oscillatory Shear Rheometry. \emph{Macromolecules} \textbf{2017}, \emph{50},
  5146--5154
\mciteBstWouldAddEndPuncttrue
\mciteSetBstMidEndSepPunct{\mcitedefaultmidpunct}
{\mcitedefaultendpunct}{\mcitedefaultseppunct}\relax
\EndOfBibitem
\bibitem[Qian \latin{et~al.}(2019)Qian, Cao, Galuska, Zhang, Xu, and
  Gu]{Qian2019}
Qian,~Z.; Cao,~Z.; Galuska,~L.; Zhang,~S.; Xu,~J.; Gu,~X. Glass Transition
  Phenomenon for Conjugated Polymers. \emph{Macromolecular Chemistry and
  Physics} \textbf{2019}, \emph{220}, 1900062
\mciteBstWouldAddEndPuncttrue
\mciteSetBstMidEndSepPunct{\mcitedefaultmidpunct}
{\mcitedefaultendpunct}{\mcitedefaultseppunct}\relax
\EndOfBibitem
\bibitem[Not()]{Note-2}
An accurate determination of glass transition temperatures, T$_g$, is a key
  factor in the study of the thermal-based properties of conjugated polymers.
  Integrating a machine learning strategy within the experimental framework to
  measure T$_g$ could help achieving efficiently this
  goal~\cite{Alesadi2022}.\relax
\mciteBstWouldAddEndPunctfalse
\mciteSetBstMidEndSepPunct{\mcitedefaultmidpunct}
{}{\mcitedefaultseppunct}\relax
\EndOfBibitem
\bibitem[Richter \latin{et~al.}(2005)Richter, Monkenbusch, Arbe, and
  Colmenero]{Richter2005}
Richter,~D.; Monkenbusch,~M.; Arbe,~A.; Colmenero,~J. \emph{Neutron Spin Echo
  in Polymer Systems}; Springer Berlin Heidelberg: Berlin, Heidelberg, 2005; pp
  67--78
\mciteBstWouldAddEndPuncttrue
\mciteSetBstMidEndSepPunct{\mcitedefaultmidpunct}
{\mcitedefaultendpunct}{\mcitedefaultseppunct}\relax
\EndOfBibitem
\bibitem[Chaney \latin{et~al.}(2022)Chaney, Levin, Schneider, and
  Toney]{Chaney2022}
Chaney,~T.~P.; Levin,~A.~J.; Schneider,~S.~A.; Toney,~M.~F. Scattering
  techniques for mixed donor–acceptor characterization in organic
  photovoltaics. \emph{Mater. Horiz.} \textbf{2022}, \emph{9}, 43--60
\mciteBstWouldAddEndPuncttrue
\mciteSetBstMidEndSepPunct{\mcitedefaultmidpunct}
{\mcitedefaultendpunct}{\mcitedefaultseppunct}\relax
\EndOfBibitem
\bibitem[Huang \latin{et~al.}(2020)Huang, Liu, Ding, and Forrest]{Huang2019}
Huang,~X.; Liu,~X.; Ding,~K.; Forrest,~S.~R. Is there such a thing as a
  molecular organic alloy? \emph{Mater. Horiz.} \textbf{2020}, \emph{7},
  244--251
\mciteBstWouldAddEndPuncttrue
\mciteSetBstMidEndSepPunct{\mcitedefaultmidpunct}
{\mcitedefaultendpunct}{\mcitedefaultseppunct}\relax
\EndOfBibitem
\bibitem[Alesadi \latin{et~al.}(2022)Alesadi, Cao, Li, Zhang, Zhao, Gu, and
  Xia]{Alesadi2022}
Alesadi,~A.; Cao,~Z.; Li,~Z.; Zhang,~S.; Zhao,~H.; Gu,~X.; Xia,~W. Machine
  learning prediction of glass transition temperature of conjugated polymers
  from chemical structure. \emph{Cell Reports Physical Science} \textbf{2022},
  \emph{3}, 100911
\mciteBstWouldAddEndPuncttrue
\mciteSetBstMidEndSepPunct{\mcitedefaultmidpunct}
{\mcitedefaultendpunct}{\mcitedefaultseppunct}\relax
\EndOfBibitem
\end{mcitethebibliography}
\end{document}